\documentclass[twocolumn,letterpaper]{IEEEAerospaceCLS}  
\usepackage{amsmath,amssymb}
\usepackage{graphicx}
\usepackage{cite}
\usepackage{pgffor}
\usepackage{adjustbox}
\usepackage{tabularx}
\usepackage{array}
\usepackage{csvsimple}
\usepackage{url}
\usepackage{soul}
\usepackage{xcolor}
\usepackage[caption=false,font=footnotesize]{subfig}
\usepackage{tikz}
\usepackage{pgfplots}
\pgfplotsset{compat=1.17}
\newcommand\changes[1]{{\color{black}#1}}

\begin{document}
\pagestyle{plain} 
\title{Real-World LoRaWAN Performance and Propagation Modeling Using UAV, Helikite, and Vehicle-Based Measurements}

\author{%
Sergio Vargas Villar, Simran Singh, Ozgur Ozdemir, Mihail Sichitiu, \.{I}smail G\"{u}ven\c{c}\\
Department of Electrical and Computer Engineering, North Carolina State University\\
Raleigh, NC 27695 USA\\[3pt]
svargas3, ssingh28, oozdemi, mlsichit, iguvenc @ncsu.edu
}

\maketitle
\thispagestyle{plain}

\begin{abstract}
\pagestyle{empty}
This paper presents a field-based evaluation of Long Range Wide Area Network (LoRaWAN) signal propagation conducted at two locations within the Aerial Experimentation and Research Platform for Advanced Wireless (AERPAW) testbed: Lake Wheeler Field and NC State University’s Centennial Campus. Three distinct transmission platforms were deployed: a ground vehicle, a multirotor drone at 50 meters, and a helikite at a steady altitude of 150 meters and 300 meters approximately. These platforms enabled a comparative study on how altitude, mobility, and terrain influence wireless signal reception across a LoRaWAN gateway network. We analyze received signal strength (RSSI) and signal-to-noise ratio (SNR) as functions of distance and spreading factor (SF). Three complementary metrics are visualized: SNR versus distance with demodulation thresholds, probability of successful reception, and SNR boxplots grouped by distance bins. These plots reveal link degradation patterns and demonstrate the role of adaptive SF selection in maintaining communication reliability.

To characterize propagation behavior, we apply a log-distance path loss model to empirical data from the ground vehicle experiment, which encompass both rural and urban non-line-of-sight (NLOS) conditions. Model parameters are optimized through error minimization techniques. Our results show that the helikite platform, due to its stable high-altitude position, provided the most reliable and consistent link performance. Conversely, the drone and vehicle exhibited higher variability due to movement, obstructions, and terrain-induced multipath. These findings demonstrate the influence of platform dynamics and altitude on LoRaWAN reception performance, providing support for future aerial network planning efforts.

\pagestyle{empty}
\textit{Index~Terms}--- Internet of Things (IoT), LoRa, LPWAN, path loss model, signal coverage analysis, spreading factor (SF), UAV.
\end{abstract}

\vspace{-0.3cm}

\tableofcontents
\section{Introduction}\label{sec:intro}

Long-Range (LoRa) is a wireless communication technology developed by Semtech that enables low-power, long-range, secure data transmission for Internet of Things (IoT) applications~\cite{SemtechCorporation2024WhatLoRa}. It utilizes chirp spread spectrum (CSS) modulation and operates in unlicensed frequency bands, often in the sub-GHz range, offering robust connectivity across large geographical areas with minimal power consumption. While the physical layer LoRa is proprietary, the LoRaWAN protocol built on top of it is open and maintained by the LoRa Alliance~\cite{SemtechCorporation2024WhatLoRa}.

\sloppy
LoRaWAN is a low-power wide area network (LPWAN) protocol that enables the connection of battery-operated IoT devices over long distances using unlicensed frequency bands, while ensuring bidirectional communication, end-to-end security, mobility, and geolocation services~\cite{SemtechCorporation2024WhatLoRaWAN}. Recognized by the International Telecommunication Union (ITU) as a global LPWAN standard, LoRaWAN provides seamless interoperability across manufacturers and supports scalable deployments for IoT applications~\cite{SemtechCorporation2024WhatLoRaWAN}. 

The Aerial Experimentation and Research Platform for Advanced Wireless (AERPAW) is a Platform for Advanced Wireless Research (PAWR) project funded by the National Science Foundation (NSF) that enables advanced wireless research through aerial and ground-based network experimentation~\cite{AERPAW2024AERPAW}. Located in Raleigh, North Carolina, AERPAW provides a testbed for studying wireless communication technologies, including 5G, IoT, and unmanned aerial systems (UAS)~\cite{AERPAW2024AERPAW}. As part of its infrastructure, AERPAW has implemented a LoRaWAN network to support low-power, large-scale IoT experiments, enabling research on long-range wireless connectivity for smart cities, environmental monitoring, and mobility applications~\cite{AERPAW2024AERPAW}.


This study investigates the real-world performance of a deployed LoRaWAN network within the AERPAW testbed at NC State, specifically at Lake Wheeler Field and Centennial Campus. The analysis focuses on key network metrics, including RSSI, SNR, SF, across different gateways and transmission scenarios.

Beyond traditional path loss estimation, the study also explores the influence of SF settings on reception probability and demodulation thresholds, providing a more complete picture of LoRaWAN behavior in practice. The results offer insights into how factors such as altitude, mobility, and SF configuration affect coverage and reliability, helping guide the planning and tuning of LoRaWAN networks for long-range IoT applications.
\begin{figure*}[htbp]
    \centering
    \subfloat[AERPAW SAM platform equipped with LoRaWAN LoStik transmitter (SPN).]{%
        \includegraphics[width=0.45\textwidth]{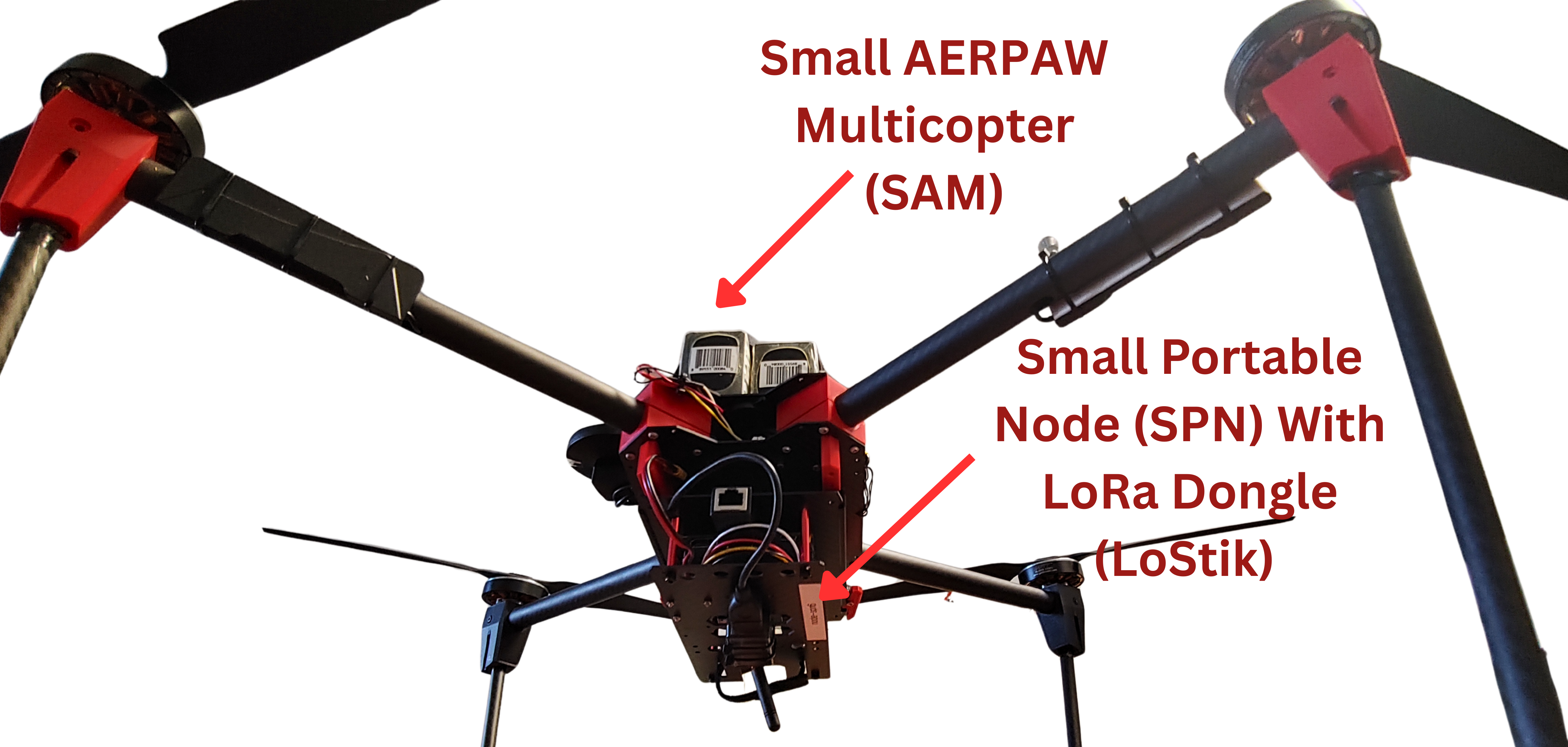}%
        \label{fig:Drone}%
    }
    \hspace{0.05\textwidth}%
    \subfloat[LoRaWAN device attached to a helikite for high-altitude stationary measurements.]{%
        \includegraphics[width=0.45\textwidth]{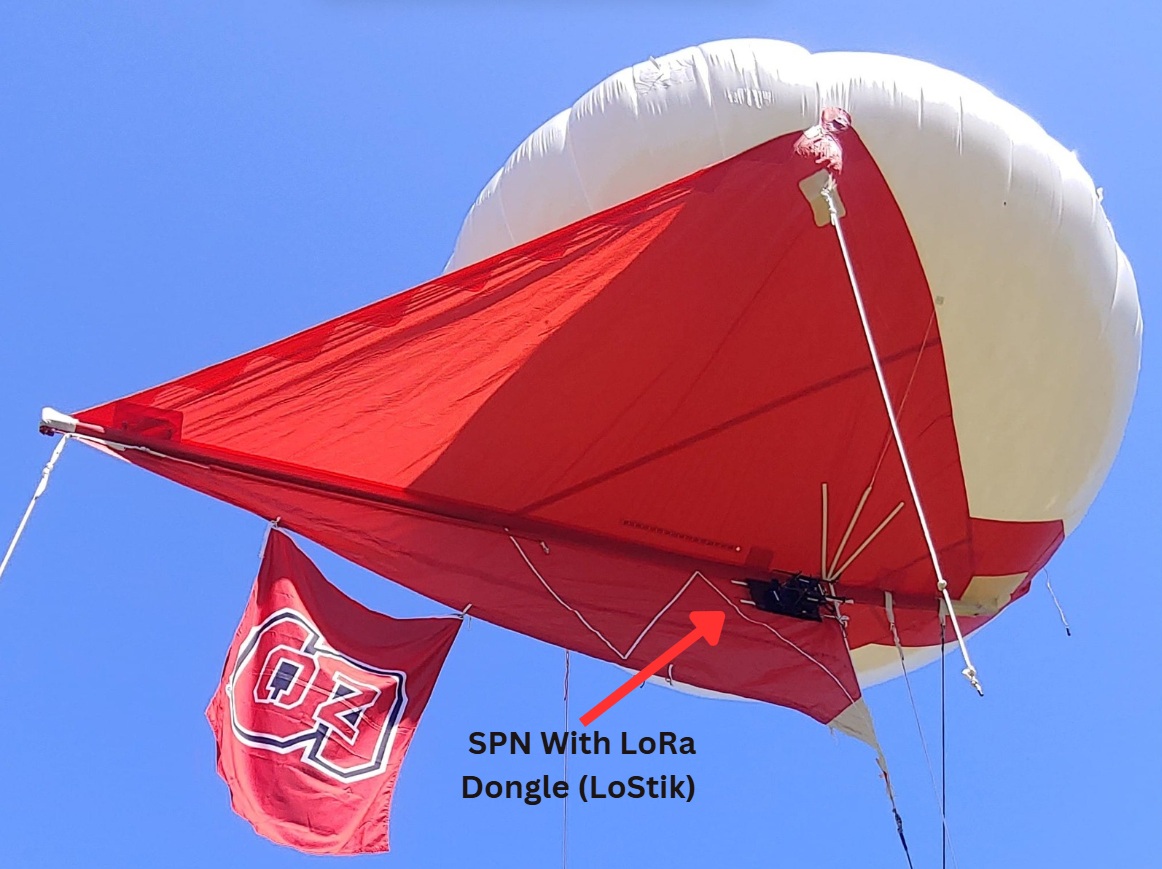}%
        \label{fig:Helikate}%
    }
    \caption{AERPAW vehicle platforms used in the experiments: small AERPAW multicopter (SAM) and helikite.}
    \label{fig:AerialPlatforms}
\end{figure*}

 \begin{figure}[!t]
     \centering
     \includegraphics[width=0.7\linewidth]{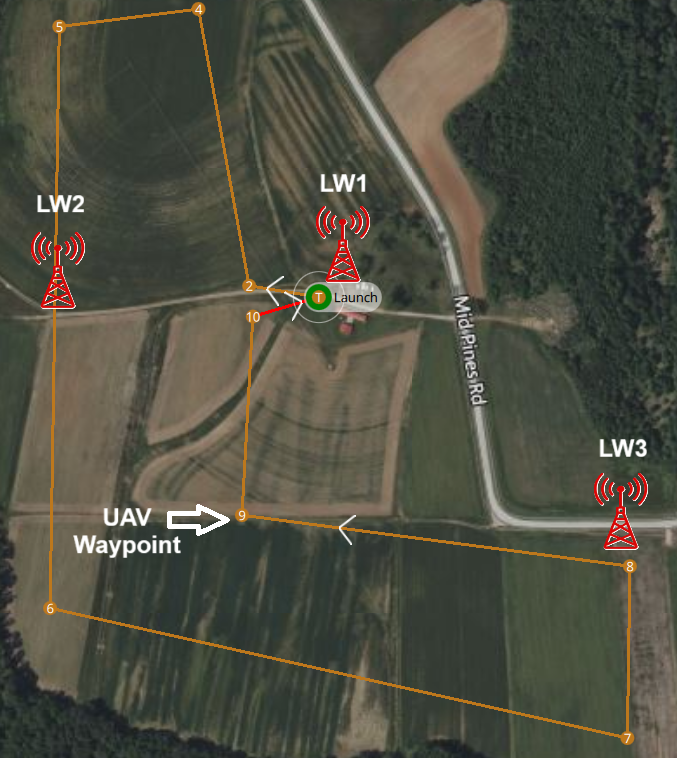}
     \caption{Predefined drone flight path used during LoRaWAN measurements at Lake Wheeler Field (open and semi-rural environment).}
     \label{fig:FlightPlanLW}
 \end{figure}


The remainder of this paper is organized as follows. Section~\ref{sec:experimental_setup} presents the experimental setup, describing the deployment locations, transmission platforms, and testbed configuration. Section~\ref{sec:system} outlines the system setup, including transmission parameters, LoRaWAN hardware, and data collection methodology. Section~\ref{sec:related_work} reviews prior research on LoRaWAN signal modeling, with a focus on empirical path loss models used in urban and rural deployments. Section~\ref{sec:propagation_model} introduces the log-distance path loss model and explains its application to the car-based measurements, along with distance calculation methods and demodulation threshold references. Section~\ref{sec:results} details the experimental findings, comparing results from three distinct test campaigns—car, helikite, and drone. It highlights variations in RSSI, SNR, and reception behavior across platforms, supported by cumulative distribution plots, SNR threshold analysis, probability-of-reception curves, and SNR boxplots grouped by spreading factor and distance.

Finally, Section~\ref{sec:conclusion} summarizes the key insights and outlines future directions for scenario-specific LoRaWAN optimization based on empirical data.

 \section{Experimental Setup and Test Areas}\label{sec:experimental_setup}

 To evaluate the real-world performance of LoRaWAN under diverse mobility and propagation environments, multiple field experiments were conducted at the AERPAW testbed. The experiments utilized different transmission platforms, including both aerial and ground-based setups, to characterize how varying deployment conditions influence LoRaWAN connectivity. The test campaigns include:

\begin{itemize}

    \item \textbf{Car-based experiment:} LoRaWAN transmissions were recorded while a vehicle equipped with a LoRa transmitter moved through different locations, including  Centennial Campus and Lake Wheeler (see Figures~\ref{fig:CC} and \ref{fig:LW}).

    \item \textbf{Helikite-based experiments:} Two separate experiments were performed using a helikite (Figure~\ref{fig:Helikate}): one at Lake Wheeler and another during the Packapalooza 2024 event, which took place on NC State’s Main Campus. The altitude was initially adjusted and then kept constant, with minor lateral drift caused by wind. Transmissions from both flights were received by gateways located at Centennial Campus and Lake Wheeler (Figures~\ref{fig:CC} and \ref{fig:LW}).

    \item \textbf{Drone-based experiment:} Conducted at Lake Wheeler, where a LoRaWAN transmitter was mounted on a drone (Figure~\ref{fig:Drone}) to evaluate aerial coverage at a fixed altitude, following a preplanned trajectory shown in Figure~\ref{fig:FlightPlanLW} (see Figure~\ref{fig:LW} for the gateway deployment).

\end{itemize}

\subsection{Experimental Test Areas}\label{sec:experimental_test_areas}

The experiment includes three main work areas shown in Figure~\ref{fig:AreasMaps}c : NC State’s Centennial and Main Campus, and Lake Wheeler Field. The car-based experiments were conducted at Centennial Campus and Lake Wheeler Field, the drone experiments were only conducted at Lake Wheeler Field, and the helikite aerial experiment was hosted on both the Main Campus and Lake Wheeler Field. The Main Campus was chosen because it represents an urban environment, Lake Wheeler Field represents a rural line-of-sight (LOS) environment, and Centennial Campus is a semi-urban area with NLOS due to buildings. Figures~\ref{fig:CC} and \ref{fig:LW} show the work areas at Centennial Campus and Lake Wheeler Field. Figure~\ref{fig:MAP} shows all three work areas marked on a map.

 \section{System Setup}\label{sec:system}

 \begin{figure*}[htbp]
     \centering
     \subfloat[]{\includegraphics[width=0.331\textwidth]{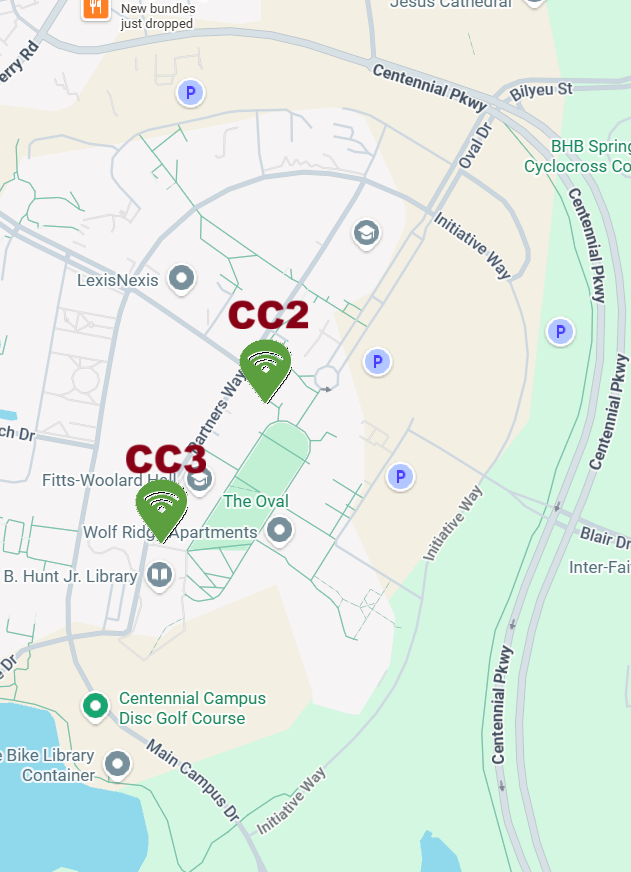}\label{fig:CC}}
     \hspace{0.02\textwidth} 
     \subfloat[]{\includegraphics[width=0.323\textwidth]{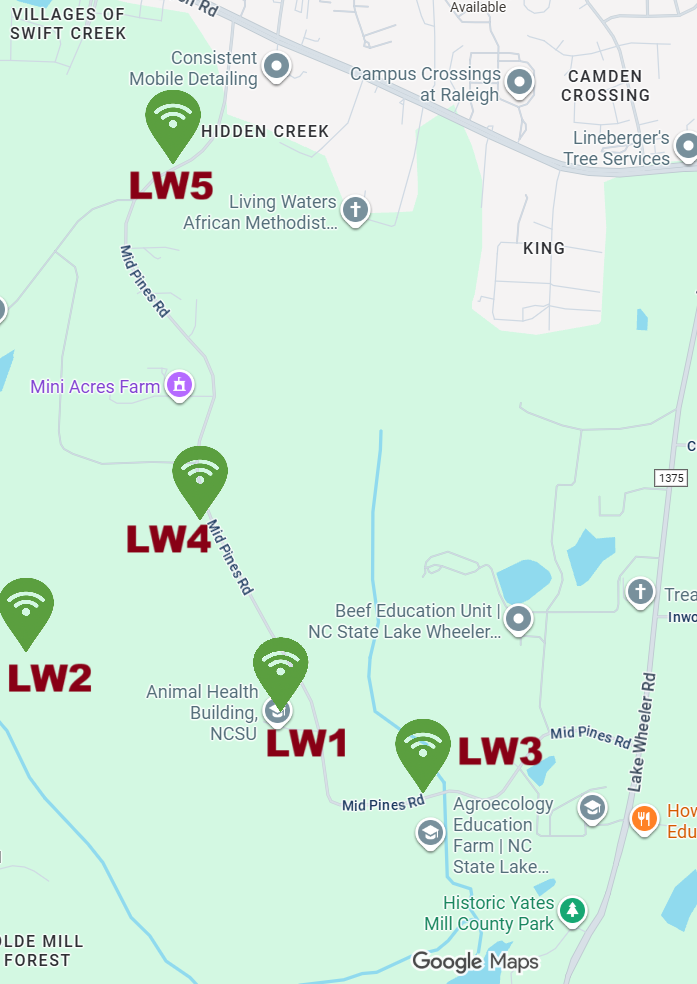}\label{fig:LW}}
      \hspace{0.02\textwidth}
     \subfloat[]{\includegraphics[width=0.27\textwidth]{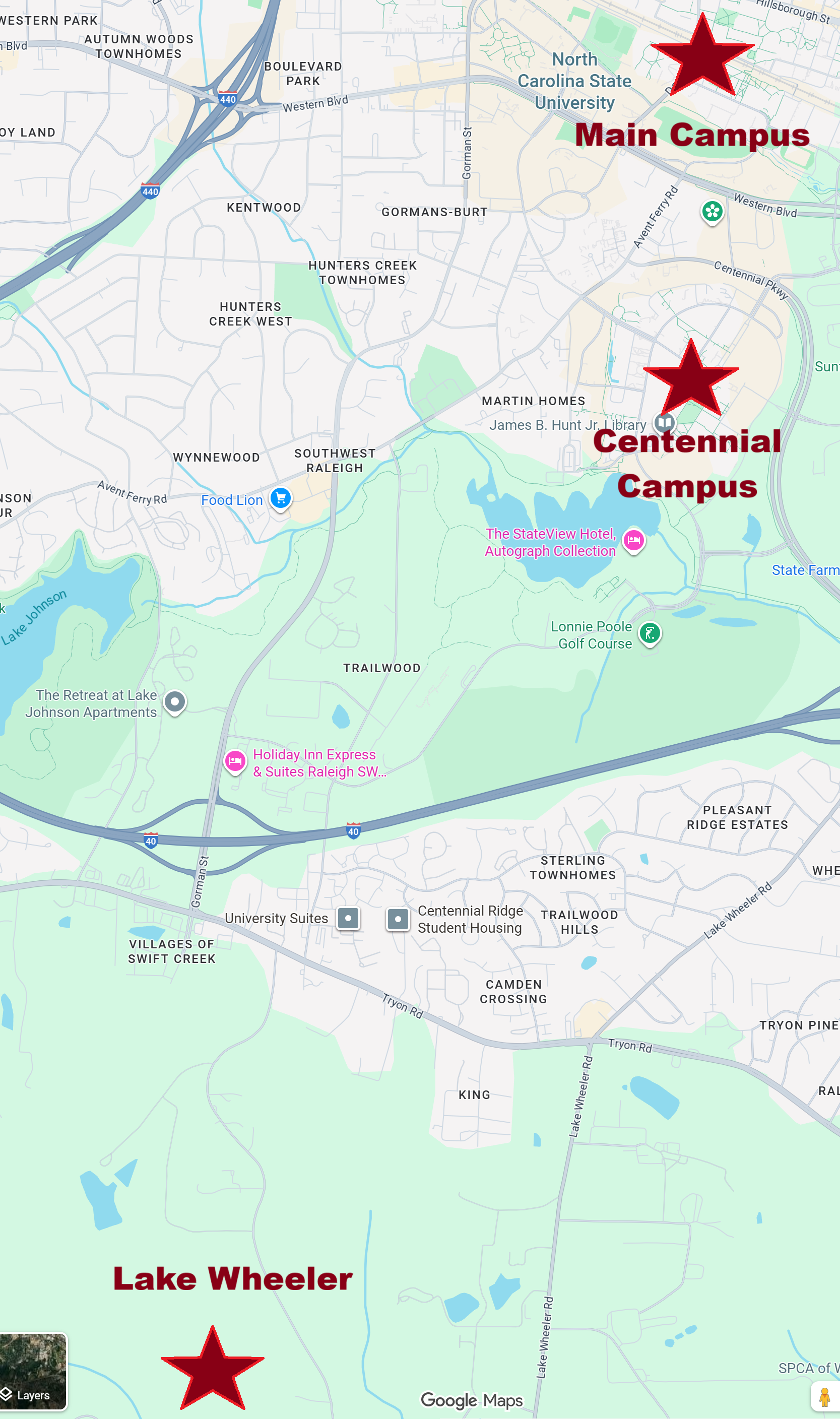}\label{fig:MAP}}
     \caption{Lab experiment areas and maps:
     (a) Deployment of LoRaWAN gateways CC2 and CC3 in NC State’s Centennial Campus (urban environment),
     (b) deployment of gateways LW1–LW5 in Lake Wheeler Field (open and semi-rural environment),
     and (c) complete map of the work area.}
     \label{fig:AreasMaps}
 \end{figure*}

For all experiments, a LoStik LoRaWAN USB transmitter by Ronoth, equipped with the Microchip RN2903 module, was used as the transmitting device. The network infrastructure consisted of seven RAK7289 LoRaWAN gateways deployed at different locations within Centennial Campus and Lake Wheeler Field, as illustrated in Figure~\ref{fig:CC} and \ref{fig:LW}. The Centennial Campus gateways, labeled CC2 and CC3, are positioned at high points on buildings to maximize coverage in the urban environment. At Lake Wheeler Field, gateways LW1 through LW5 are distributed across the open and semi-rural site to ensure broad area coverage and effective line-of-sight communication for aerial and ground experiments.

The transmitter device recorded real-time information such as geographic location, transmission timestamps, and assigned data rates (DR). On the receiving side, data logged included RSSI, SNR, reception channel, received timestamp, operating frequency, spreading factor (SF), radio frequency (RF) chain, and bandwidth.

To maintain consistency across experiments, the transmitted data packets were fixed to 7-byte hexadecimal values containing a unique packet number and timestamp. Transmissions were performed using manually selected data rates (DR0 to DR3), corresponding to spreading factors SF10 through SF7, respectively, with a fixed inter-packet delay of 2.5~seconds to ensure a controlled and repeatable measurement process. The LoStik transmitter operated in compliance with the LoRaWAN media access control (MAC) layer, and the data rate settings remained constant throughout each measurement scenario.

The transmission syntax followed the RN2903 module's command structure, as described in the RN2903 LoRa\texttrademark{} Technology Module Command Reference User’s Guide~\cite{Microchip2018RN2903_Reference}:

\begin{verbatim}
mac tx <type> <portno> <data>
\end{verbatim}
\noindent
where:
\begin{itemize}
    \item \texttt{<type>} specifies the uplink payload type: either \texttt{cnf} (confirmed) or \texttt{uncnf} (unconfirmed),
    \item \texttt{<portno>} represents the port number (1 to 223),
    \item \texttt{<data>} contains the hexadecimal payload, with length constraints dependent on the data rate.
\end{itemize}

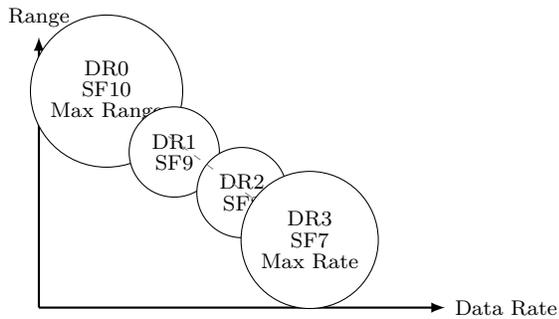
\begin{figure}[htbp]
    \centering
    \begin{tikzpicture}[>=latex, scale=0.9]

        \draw[->, thick] (0,0) -- (6,0) node[right]{\footnotesize \textbf{Data Rate}};
        \draw[->, thick] (0,0) -- (0,4) node[above]{\footnotesize \textbf{Range}};

        \node[circle,draw,fill=white,minimum size=1.2cm,align=center,font=\scriptsize] 
            (dr0) at (1,3.2) {DR0\\SF10\\\textit{Max Range}};
        \node[circle,draw,fill=white,minimum size=1.2cm,align=center,font=\scriptsize] 
            (dr1) at (2,2.3) {DR1\\SF9};
        \node[circle,draw,fill=white,minimum size=1.2cm,align=center,font=\scriptsize] 
            (dr2) at (3,1.7) {DR2\\SF8};
        \node[circle,draw,fill=white,minimum size=1.2cm,align=center,font=\scriptsize] 
            (dr3) at (4,1) {DR3\\SF7\\\textit{Max Rate}};

        \draw[dashed, gray] (dr0) -- (dr3);

    \end{tikzpicture}
    \caption{Trade-off between range and data rate for LoRa DR/SF combinations (125 kHz).}
    \label{fig:lora_dr_tradeoff}
\end{figure}

In LoRaWAN, DR defines the spreading factor (SF) and bandwidth used for transmissions, according to the regional specifications established by the LoRa Alliance~\cite{LoRaAlliance2021LoRaWANSpecification}. For the US915 frequency band, which was used in our experiments, DR0 corresponds to SF10 with a bandwidth of 125~kHz, DR1 corresponds to SF9 at 125~kHz, DR2 corresponds to SF8 at 125~kHz, and DR3 corresponds to SF7 at 125~kHz. These combinations directly affect the network’s range and throughput: lower data rates (higher SF, e.g., DR0/SF10) allow signals to propagate farther but at lower data speeds, while higher data rates (lower SF, e.g., DR3/SF7) provide faster transmission rates but shorter effective range. In our experiments, specific data rates (and thus spreading factors) were manually selected and kept constant throughout each measurement scenario. As shown in Figure \ref{fig:lora_dr_tradeoff}, these trade-offs are clearly illustrated.

\section{Related Work}\label{sec:related_work}

LoRaWAN performance has been extensively analyzed in different propagation environments using empirical, semi-deterministic, and theoretical models. Several studies have focused on evaluating path loss models to improve LoRaWAN signal predictions in urban, suburban, and rural deployments.

Dieng et al.~\cite{Dieng2020ComparingNetworks} analyzed the performance of several empirical path loss models, including the log-distance path loss model, in LoRaWAN networks. Their study highlighted that the log-distance model offers a flexible approach to predicting signal attenuation by adjusting the path loss exponent $n$ based on environmental conditions, making it suitable for diverse deployment scenarios. Ingabire et al.~\cite{Ingabire2020PerformanceEnvironment} compared multiple propagation models, including ITU-R P.1225, log-distance, Okumura-Hata, and WINNER II, in an urban LoRaWAN deployment. Their results indicated that while ITU-R P.1225 performed well in certain urban conditions, the log-distance model provided comparable accuracy with the advantage of being adaptable to different terrain types. In~\cite{Harinda2020Trace-drivenScenario}, the authors performed a trace-driven simulation of LoRaWAN air channel propagation in an urban environment, evaluating models such as Deygout 94, ITU-R 525/526, and COST-Walfish Ikegami (COST-WI). Their findings emphasized the significance of refining empirical models like log-distance to enhance LoRaWAN coverage predictions. Kucherov et al.~\cite{Kucherov2021InvestigationSignals} studied the impact of multipath propagation on LoRaWAN transmissions, demonstrating how signal reflections and environmental obstacles affect RSSI and path loss. Their results suggest that LoRa’s chirp spread spectrum (CSS) modulation improves resilience against multipath fading, reinforcing the importance of environment-specific tuning of path loss models. In~\cite{Lima2024LoRaRegions}, the authors investigated LoRaWAN deployments in Amazonian regions, highlighting the need for location-specific tuning of propagation models in dense vegetation and remote environments. Their work demonstrates that models like log-distance can be adjusted for better accuracy when calibrated with real-world measurements.

\changes{Along with LoRa path loss analysis, studies in literature also focus on generating measurement datasets through campaigns conducted in urban, suburban, and rural scenarios. These studies employ  mobile or static end-nodes and record metrics such as signal to noise ratio (SNR), received signal strength indicator (RSSI), and packet received ratio (PRR). Liando et al.~\cite{liando2019known} measured PRR as a function of distance between end-nodes and gateways on a university campus characterized by high-rise buildings and vegetation, deriving coverage boundaries based on measured signal strength. Similarly, RSSI and SNR measurements were used to estimate PRR in rural and suburban campus environments in Argentina, where reception margin variations with distance were analyzed~\cite{matelica2025lora}. Vatcharatiansakul et al.~\cite{vatcharatiansakul2017experimental} evaluated LoRa SNR and packet loss as functions of distance in urban indoor and outdoor environments in Thailand, additionally examining the impact of antenna height and gain on coverage. A large-scale urban LoRaWAN deployment consisting of 1,000 gateways and 19,821 end-nodes, covering an area of $130~\mathrm{km}^2$, was presented in~\cite{tong2023citywide}, where RSSI and SNR values were reported per gateway for each uplink transmission and used to evaluate packet loss rates across twelve smart city applications. Khoury et al.~\cite{khoury2025performance} studied variations in RSSI and SNR of commercial LoRa IoT devices as a function of distance in an urban environment using a ground vehicle. Specialized datasets have also been developed for unique scenarios, including search-and-rescue operations in avalanche environments~\cite{girolami2024experimental}, air-to-water and air-to-ground propagation in mountainous regions~\cite{druagulinescu2024understanding}, and UAV-assisted LoRaWAN deployments, utilizing a gateway mounted on a hovering UAV, end-devices and multiple gateways on the ground, and a cloud server~\cite{limeasurement}.}

These studies demonstrate the importance of \changes{LoRaWAN measurement datasets,} accurate signal modeling, and the flexibility of the log-distance path loss model when adapted to real-world conditions. Building on this foundation, our work \changes{creates LoRaWAN measurement datasets for mobile aerial and ground scenarios, and} incorporates the log-distance model to characterize propagation behavior in UAV-, Helikite-, and car-based experiments, specifically capturing the non-line-of-sight (NLOS) behavior in mixed rural and urban scenarios. The broader focus of this study is on evaluating how signal behavior varies with SF and transmission conditions. By analyzing reception probability, demodulation thresholds, and SNR distributions, we provide a more comprehensive understanding of LoRaWAN performance across different platforms and deployment environments.

\newcolumntype{C}[1]{>{\centering\arraybackslash}m{#1}}
\begin{table}[!t]
\renewcommand{\arraystretch}{1.2}
\centering
\caption{Summary of LoRaWAN experimental configuration and deployment parameters.}
\label{tab:experiment_setup}
\small
\begin{tabular}{|C{3.5cm}|C{4cm}|}
\hline
\textbf{Parameter} & \textbf{Value} \\ \hline
Transmitter & LoStik RN2903 (Ronoth) \\ \hline
Gateways & 7× RAK7289 \\ \hline
SF & 7–10 \\ \hline
DR & DR0–DR3 \\ \hline
Packet interval & 2.5~s \\ \hline
Payload size & 7~bytes (hex) \\ \hline
Experiment locations & Lake Wheeler, Centennial and Main Campus \\ \hline
Experiment types & Drone-based, helikite-based, car-based \\ \hline
Mobility & Fixed (helikite), aerial (drone), ground mobile (car) \\ \hline
\end{tabular}
\end{table}

\section{Propagation Model}\label{sec:propagation_model}

Accurate propagation models are crucial for effective LoRaWAN network planning. This study focuses on optimizing the log-distance path loss model, which is widely used in LoRaWAN performance evaluations due to its adaptability to various environments~\cite{Dieng2020ComparingNetworks}.

Our approach involves fine-tuning the model’s parameters using experimental data collected on the AERPAW testbed, ensuring accurate prediction of signal attenuation in real deployment environments.

\subsection{Log-Distance Path Loss Model}

The log-distance path loss model expresses the relationship between path loss $PL$ and distance $d$ in a logarithmic form:
\begin{equation}
PL(d) = PL(d_0) + 10\,n \log_{10} \left( \frac{d}{d_0} \right) + X_\sigma,
\label{eq:logdistance}
\end{equation}
where:
\begin{itemize}
    \item $PL(d)$ is the estimated path loss at distance $d$ (in dB),
    \item $PL(d_0)$ is the reference path loss at a reference distance $d_0$,
    \item $n$ is the path loss exponent, which varies depending on the propagation environment,
    \item $d$ is the distance between the transmitter and receiver (in meters),
    \item $d_0$ is the reference distance (in meters), ensuring a minimum free-space loss constraint,
    \item $X_\sigma$ is a zero-mean Gaussian variable accounting for shadow fading effects~\cite{Dieng2020ComparingNetworks}.
\end{itemize}

Unlike deterministic models such as ITU-R P.1225 or Okumura-Hata, the log-distance model allows flexibility in tuning parameters $n$ and $PL(d_0)$ to better match real-world LoRaWAN deployments~\cite{Dieng2020ComparingNetworks}. This adaptability makes it suitable for diverse propagation environments, as demonstrated in several LoRaWAN studies where environment-specific calibration significantly improved accuracy~\cite{Ingabire2020PerformanceEnvironment}.

\subsection{Geodesic Distance Calculation in 3D}

To ensure accurate path loss modeling, we compute the true 3D distance between the LoRaWAN transmitter and each receiving gateway, accounting for both surface curvature and elevation differences. The horizontal (great-circle) distance is calculated using the haversine formula:
\begin{equation}
d_{\text{horiz}} = 2 R \cdot \arctan2\left( \sqrt{a}, \sqrt{1 - a} \right),
\end{equation}
\begin{equation}
a = \sin^2\left( \frac{\Delta \phi}{2} \right) + \cos(\phi_1) \cos(\phi_2) \sin^2\left( \frac{\Delta \lambda}{2} \right),
\end{equation}
where:
\begin{itemize}
    \item $R$ is the average Earth radius, taken as 6,371~km,
    \item $\phi_1$, $\phi_2$ are the latitudes (in radians) of the transmitter and receiver,
    \item $\Delta \phi$ and $\Delta \lambda$ represent the differences in latitude and longitude, respectively, expressed in radians.
\end{itemize}

To incorporate vertical separation, we use the transmitter and gateway altitudes, denoted as $h_1$ and $h_2$, and compute the total 3D distance as:
\begin{equation}
d_{\text{3D}} = \sqrt{d_{\text{horiz}}^2 + (h_2 - h_1)^2}.
\end{equation}

This formulation ensures that gateways located at different elevations are correctly accounted for in the path loss model. The implementation is based on methods described by Veness in~\cite{ChrisVeness2020MovableDistance}.

\begin{figure}[!t]
    \centering
    \includegraphics[width=0.8\linewidth]{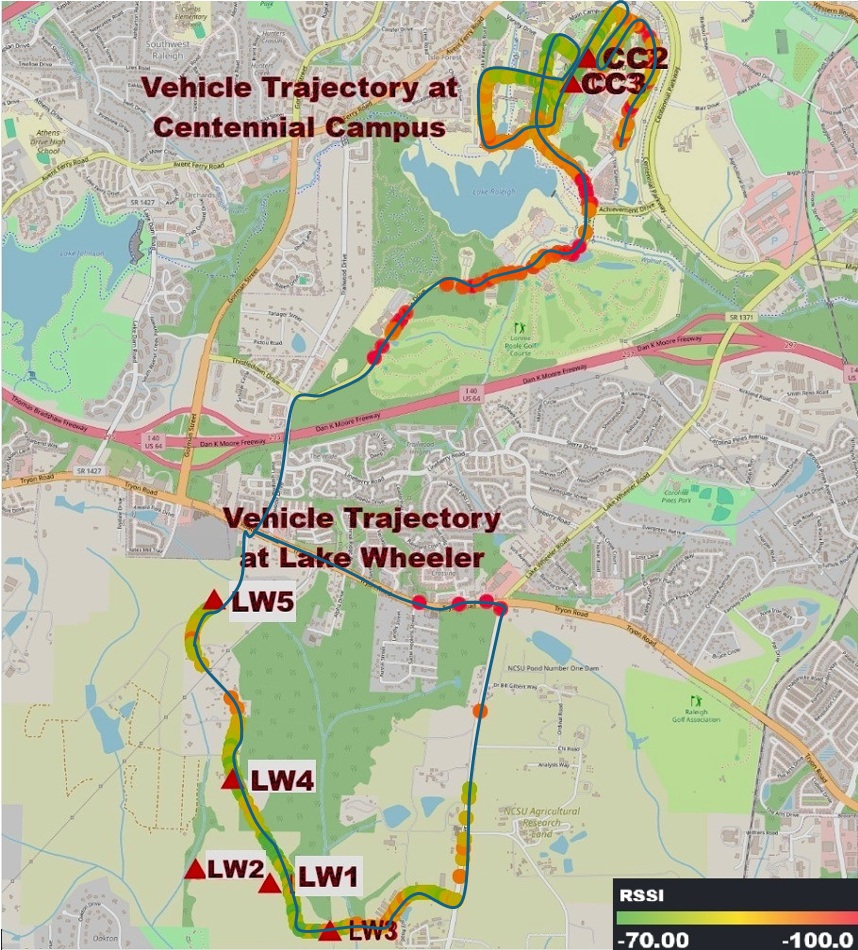}
    \caption{Vehicle trajectory and gateway locations with RSSI overlay – car-based experiment.}
    \label{fig:car_rssi_map}
\end{figure}

\subsection{Application to LoRaWAN Measurements}

To evaluate the applicability of the log-distance path loss model, we apply it to the car-based experiment conducted in the AERPAW testbed. In this setup, a LoStik LoRaWAN transmitter sends packets at manually selected data rates corresponding to specific SF values (SF7 to SF10). The packets are received by multiple RAK7289 gateways deployed at fixed locations across the test site.

The measured path loss for each received packet is computed as:
\begin{equation}
PL_{\text{measured}} = P_{\text{tx}} - \text{RSSI},
\end{equation}
where:
\begin{itemize}
    \item $P_{\text{tx}}$ is the transmit power, fixed at 20~dBm,
    \item RSSI is the received signal strength indicator measured at the gateway.
\end{itemize}

To model the observed signal attenuation, we fit the following log-distance path loss equation using linear regression in the log-distance domain:
\begin{equation}
PL(d) = A + 10\,n \log_{10}(d),
\end{equation}
where $A$ is the intercept and $n$ is the path loss exponent. These parameters are estimated using all valid packet receptions from the car platform. This approach captures signal attenuation trends under realistic rural and urban NLOS conditions using a single unified model, rather than applying per-gateway calibration. While prior studies emphasize the need for scenario-specific path loss calibration due to varying urban density and mobility~\cite{Dieng2020ComparingNetworks, Ingabire2020PerformanceEnvironment}, our study adopts a simplified approach by fitting a single path loss model for the car experiment. This reflects a generalized NLOS environment that combines rural and urban propagation effects.

\begin{figure*}[htbp]
    \centering
    \subfloat[CDF of RSSI across gateways]{\includegraphics[width=0.45\textwidth]{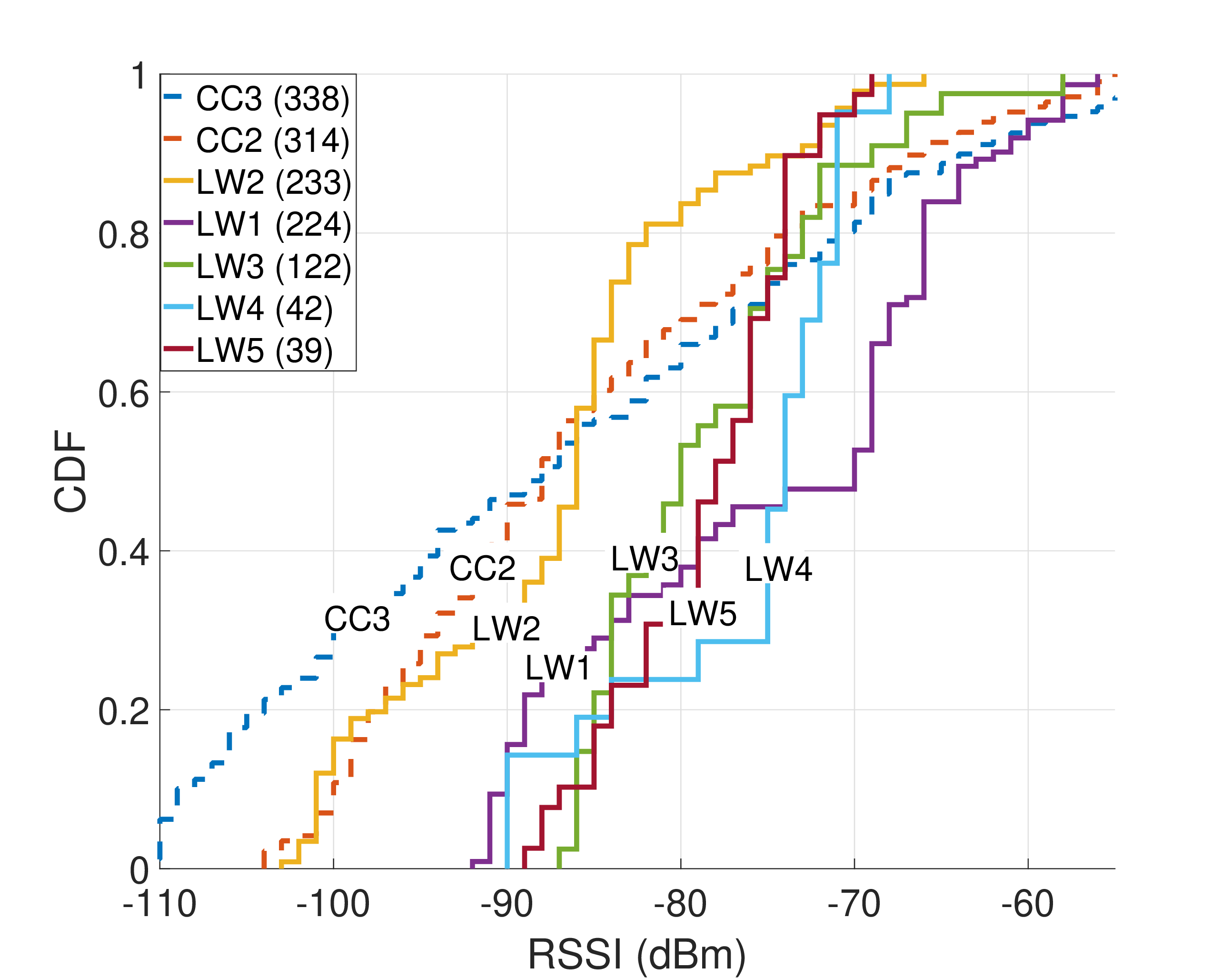}\label{fig:rssi_cdf}}
    \hfill
    \subfloat[CDF of SNR across gateways]{\includegraphics[width=0.45\textwidth]{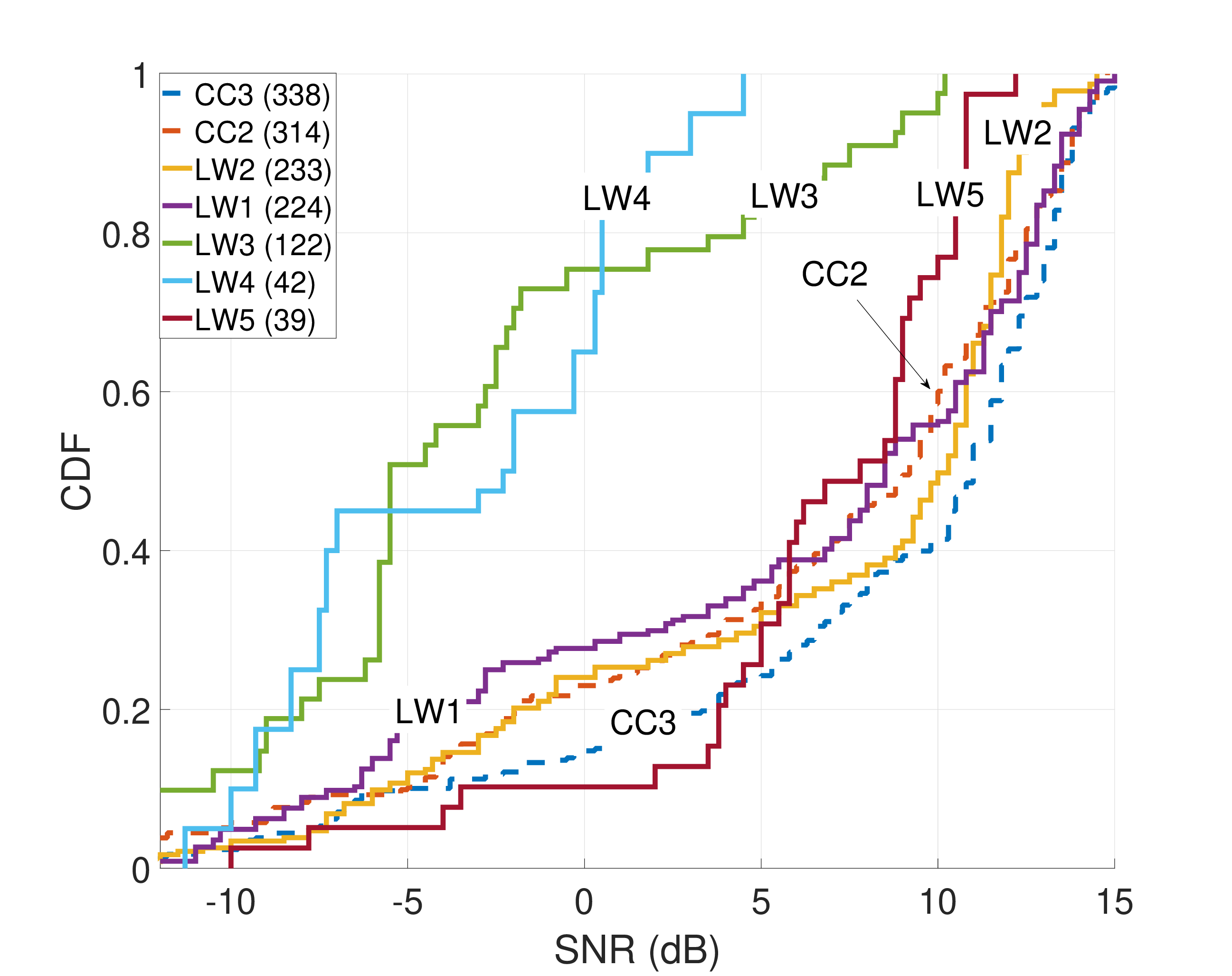}\label{fig:snr_cdf}}
    \par\bigskip
    \subfloat[SNR vs distance by SF]{\includegraphics[width=0.45\textwidth]{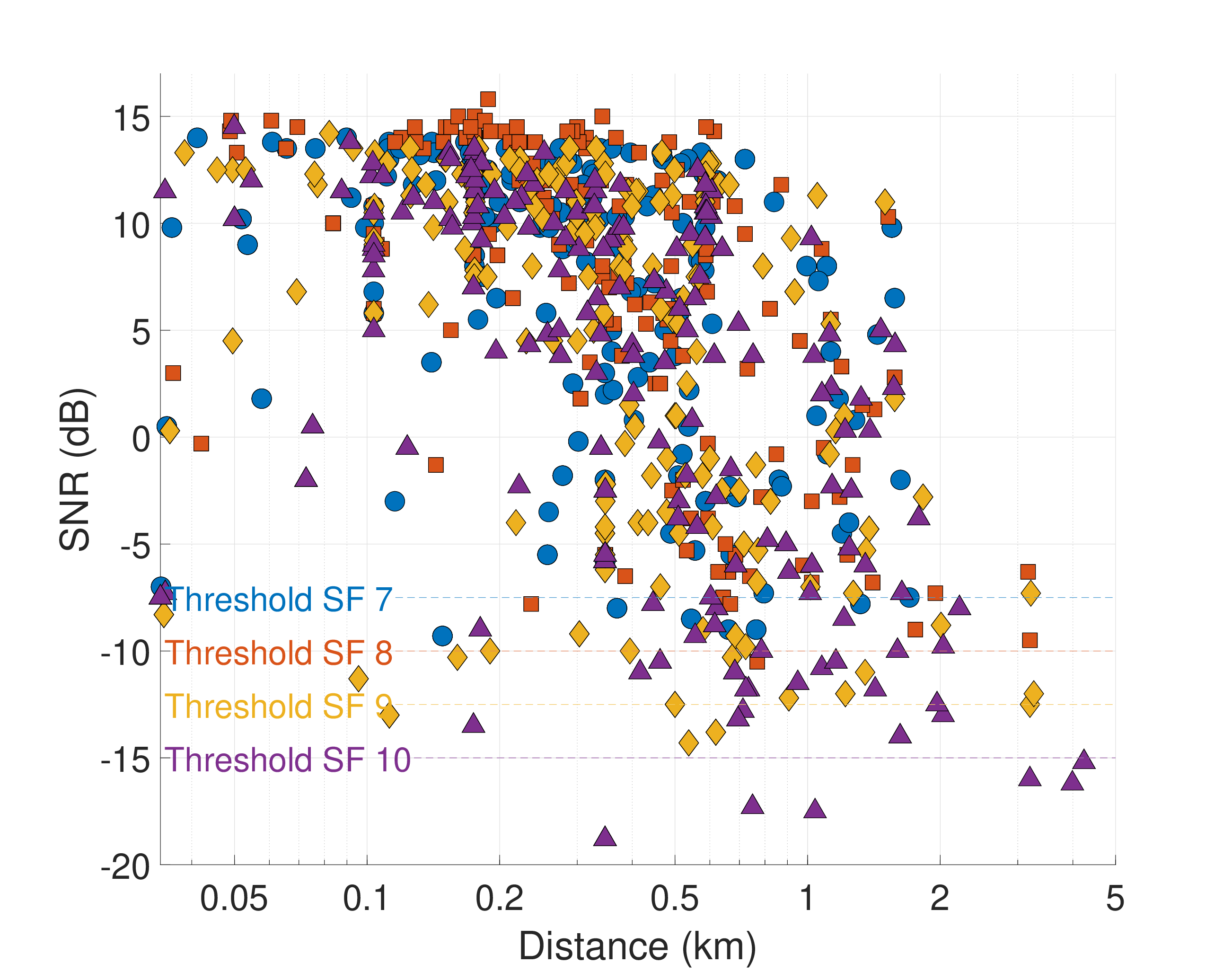}\label{fig:snr_vs_dist}}
    \hfill
    \subfloat[Probability of reception by SF]{\includegraphics[width=0.45\textwidth]{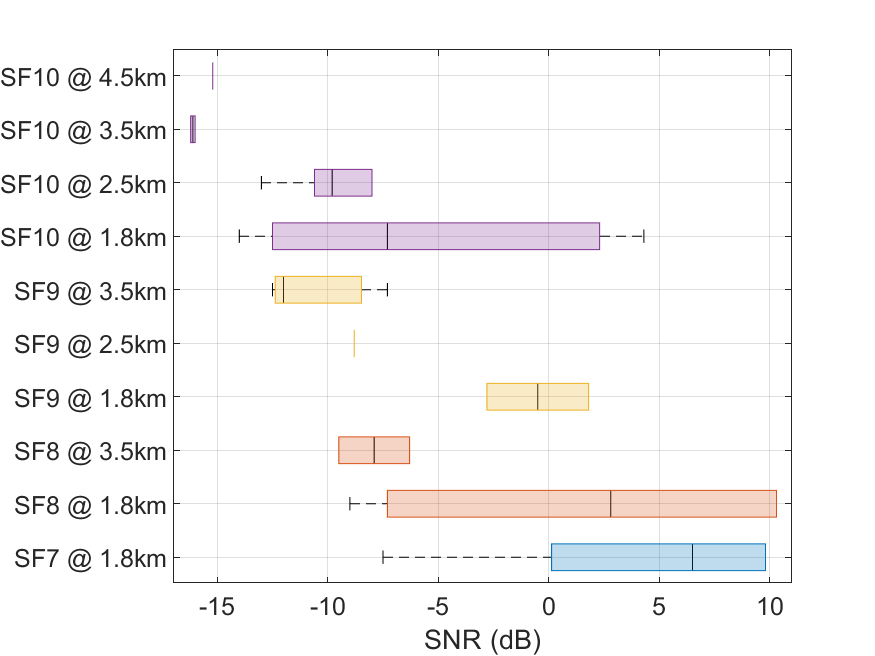}\label{fig:snr_box}}
    \par\bigskip
    \subfloat[Boxplot of SNR vs distance by SF]{\includegraphics[width=0.45\textwidth]{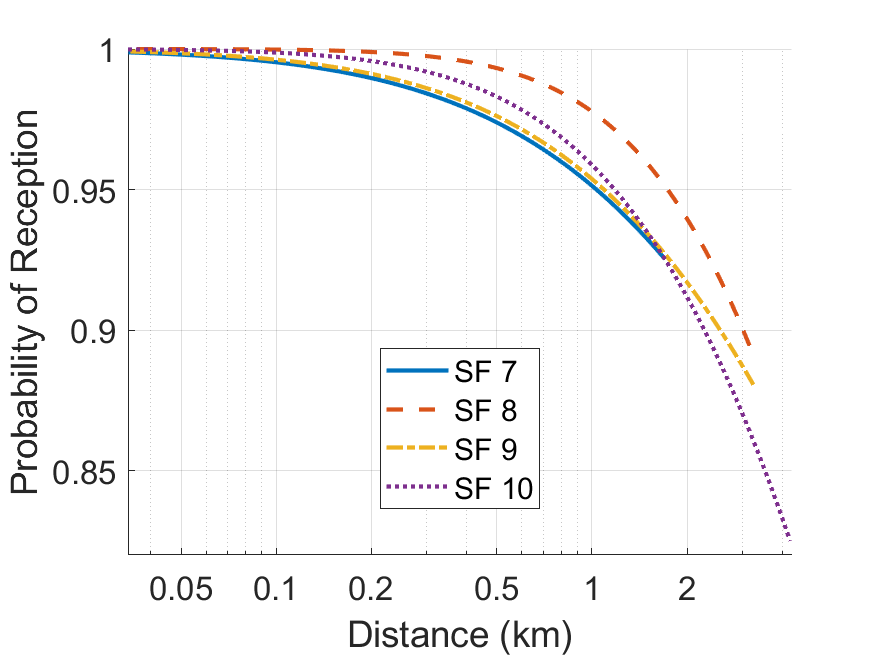}\label{fig:recv_prob}}
    \hfill
    \subfloat[]{\includegraphics[width=0.45\textwidth]{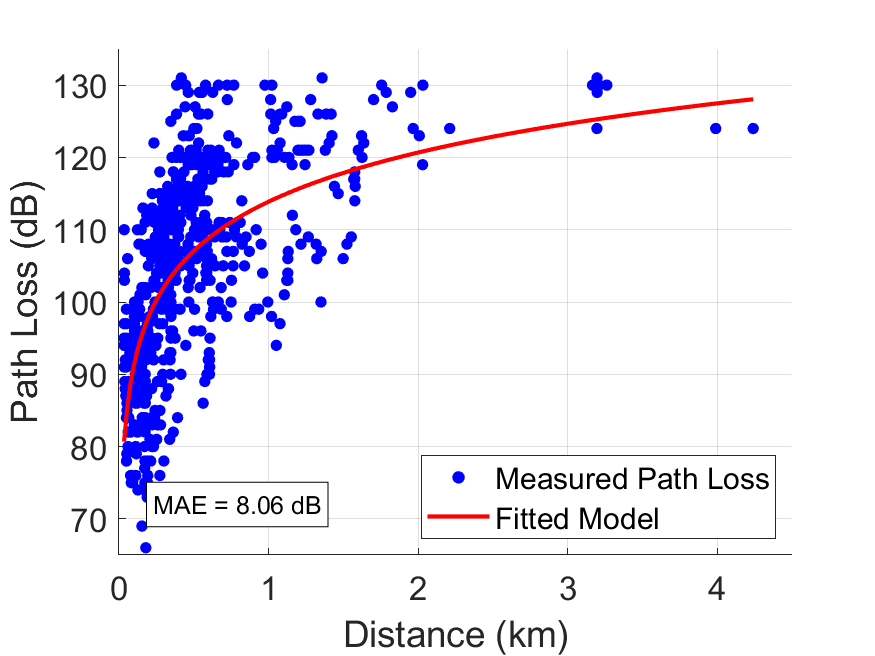}\label{fig:pathloss}}
    \caption{Car experiment results: 
        (a) RSSI CDF per gateway, 
        (b) SNR CDF per gateway, 
        (c) SNR vs distance by spreading factor,
        (d) Boxplot of SNR vs distance by SF, 
        (e) Reception probability vs distance by SF,
        (f) Path loss model fit to car measurements.
    }
    \label{fig:car_all}
\end{figure*}

\subsection{Spreading Factor Sensitivity and Reception Thresholds}

The SF determines the number of chips per symbol used in LoRa modulation, with higher SFs resulting in longer transmission times and greater sensitivity. Each SF requires a minimum SNR level to successfully demodulate a packet. These SNR thresholds define the reception limit for a given configuration and are commonly used in link budget analyses and performance models. The values used in this study are derived from the SX1276 LoRa transceiver specifications~\cite{SemtechSemtechSX1276} and are summarized in Table~\ref{tab:snr_thresholds}.

\begin{table}[ht]
\centering
\caption{SNR thresholds required for successful demodulation at different LoRa spreading factors (based on the SX1276 transceiver specifications~\cite{SemtechSemtechSX1276}).}
\label{tab:snr_thresholds}
\begin{tabular}{|c|c|c|}
\hline
\textbf{SF} & \textbf{Chips/symbol} & \textbf{SNR threshold (dB)} \\
\hline
6 & 64 & $-5.0$ \\
7 & 128 & $-7.5$ \\
8 & 256 & $-10.0$ \\
9 & 512 & $-12.5$ \\
10 & 1024 & $-15.0$ \\
11 & 2048 & $-17.5$ \\
12 & 4096 & $-20.0$ \\
\hline
\end{tabular}
\end{table}

The SNR thresholds in Table~\ref{tab:snr_thresholds} are used as reference limits in our analysis to determine whether a received signal is likely to be successfully decoded. Specifically, for each SF, we assess whether the received SNR exceeds its respective threshold. These thresholds are visualized as horizontal dashed lines in the SNR vs. distance plots.

To further evaluate link reliability, we model the probability of successful reception as a function of distance using a fitted Gaussian distribution of SNR values. The reception probability $P_{\text{recv}}(d)$ is estimated for each SF by computing the likelihood that the received SNR exceeds the demodulation threshold $\gamma_{\text{SF}}$, given a distance-dependent mean SNR and a standard deviation $\sigma$:

\begin{equation}
    P_{\text{recv}}(d) = 1 - \Phi\left( \frac{\gamma_{\text{SF}} - \mu(d)}{\sigma} \right),
    \label{eq:probReceive}
\end{equation}

where:
\begin{itemize}
    \item $\mu(d)$ is the mean SNR estimated by fitting a linear model in $\log_{10}(d)$,
    \item $\Phi(\cdot)$ is the cumulative distribution function (CDF) of the standard normal distribution,
    \item $\gamma_{\text{SF}}$ is the SNR threshold for the respective SF.
\end{itemize}

In addition to the reception probability model, we include boxplots that show how SNR values are distributed across different SFs and distance ranges.

\section{Results and Analysis}\label{sec:results}

To evaluate the performance of LoRaWAN in real-world conditions, a series of experiments were conducted across different environments and mobility scenarios. Table~\ref{tab:experiment_setup} summarizes the key parameters used in these experiments, including transmission settings, network infrastructure, and measurement details.

\subsection{Car-Based Experiment}

The car-based experiment was conducted to evaluate LoRaWAN signal propagation under mobile, ground-level conditions. A LoStik LoRaWAN transmitter was installed on the vehicle’s roof and operated along a predefined route covering both Lake Wheeler Field and Centennial Campus. This trajectory spanned diverse environments, including open rural areas, tree-lined roads, and urban zones with building obstructions.

The goal was to capture signal behavior across a broad range of distances and propagation conditions, offering a valuable comparison point against the aerial measurements conducted with the drone and helikite. Unlike the aerial platforms, the vehicle’s low-altitude movement and ground-level proximity introduced stronger multipath propagation, increased shadowing, and greater signal variability, reflecting conditions commonly encountered in real-world mobile and urban IoT deployments.

\begin{figure*}[htbp]
    \centering
    \subfloat[Helikite trajectory and gateway locations with RSSI overlay.]
        {\includegraphics[width=0.4\textwidth]{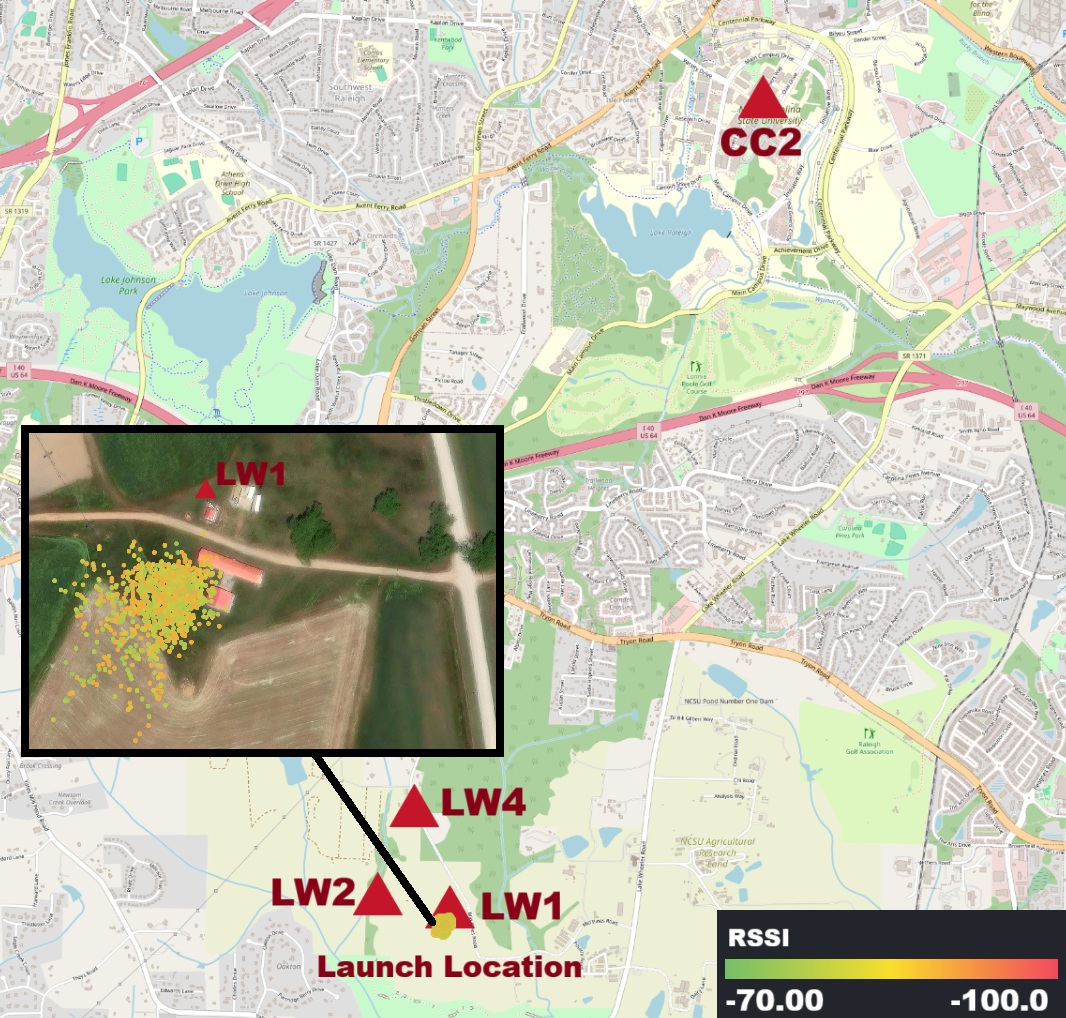}\label{fig:helikite_lw_map}}
    \hfill
    \subfloat[CDF of RSSI across gateway. CC2 is in urban environment, while others are in a rural suburban environment surrounded by trees.]
        {\includegraphics[width=0.4\textwidth]{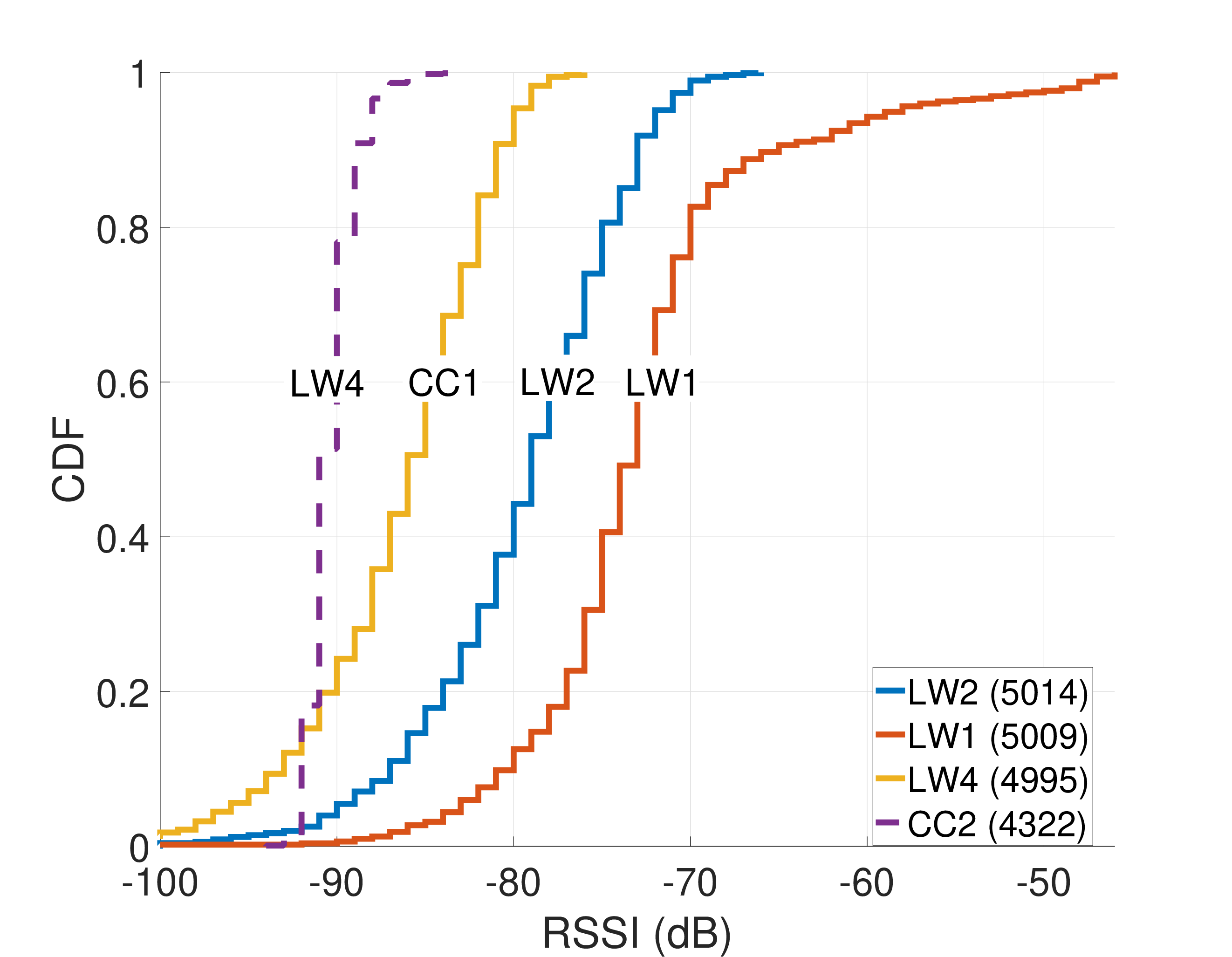}\label{fig:helikite_lw_rssi_all}}
    \par\bigskip
    \subfloat[CDF of SNR across gateways. CC2 is in urban environment, while others are in a rural suburban environment surrounded by trees.]
        {\includegraphics[width=0.4\textwidth]{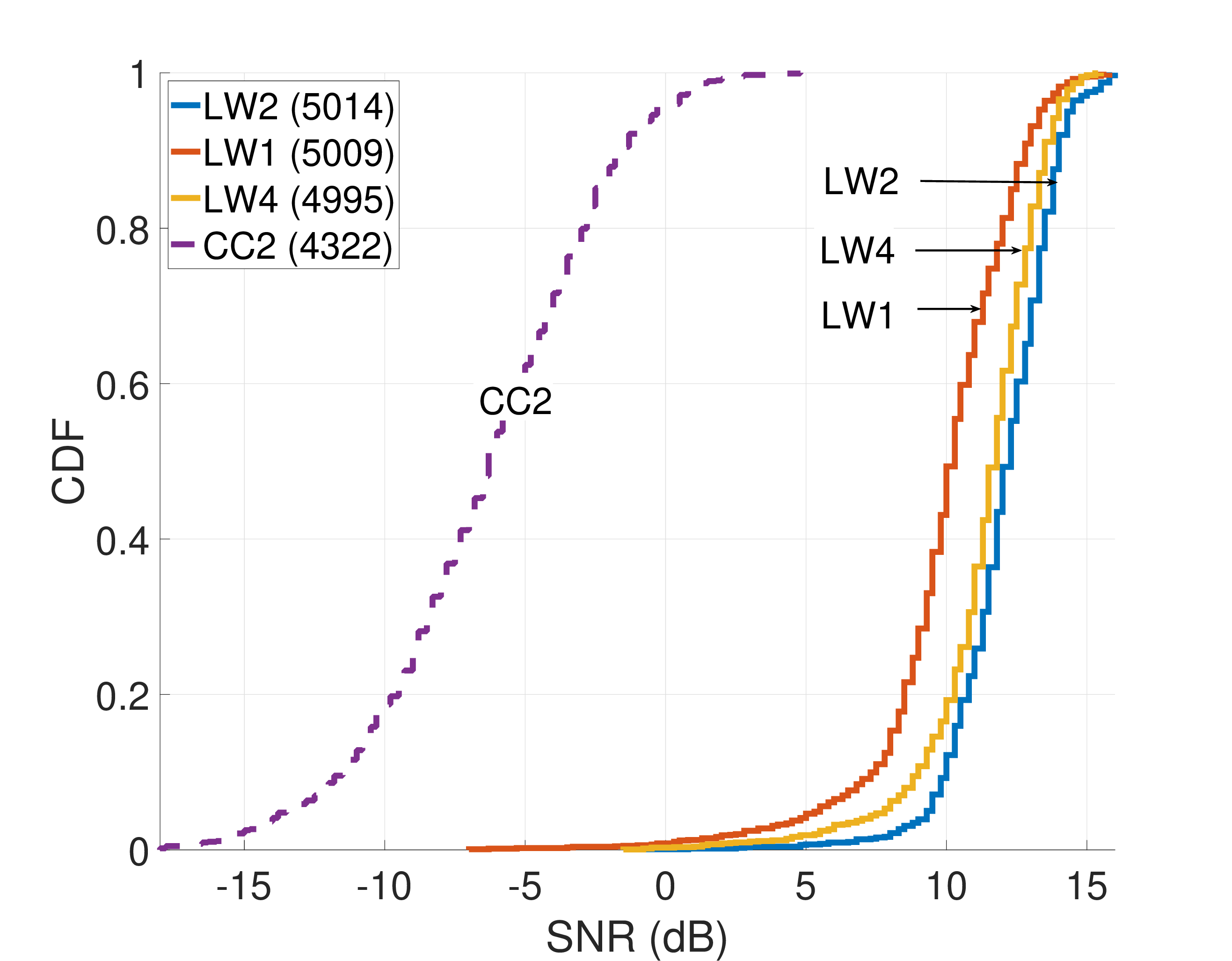}\label{fig:helikite_lw_snr_all}}
    \hfill
    \subfloat[SNR vs distance by SF.]
        {\includegraphics[width=0.4\textwidth]{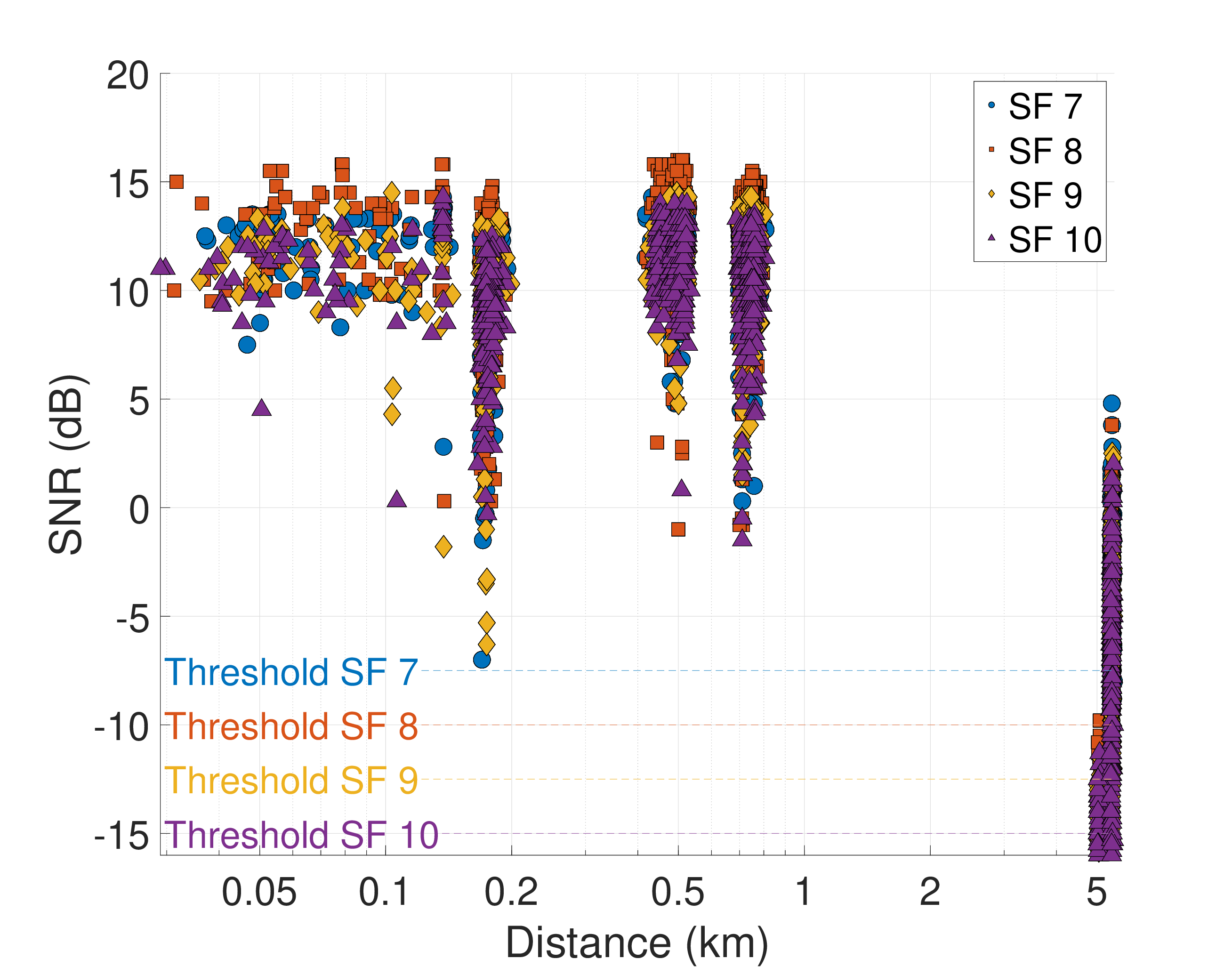}\label{fig:helikite_lw_SNR_vs_Distance_By_SF}}
    \par\bigskip
        \subfloat[Probability of reception.]
        {\includegraphics[width=0.4\textwidth]{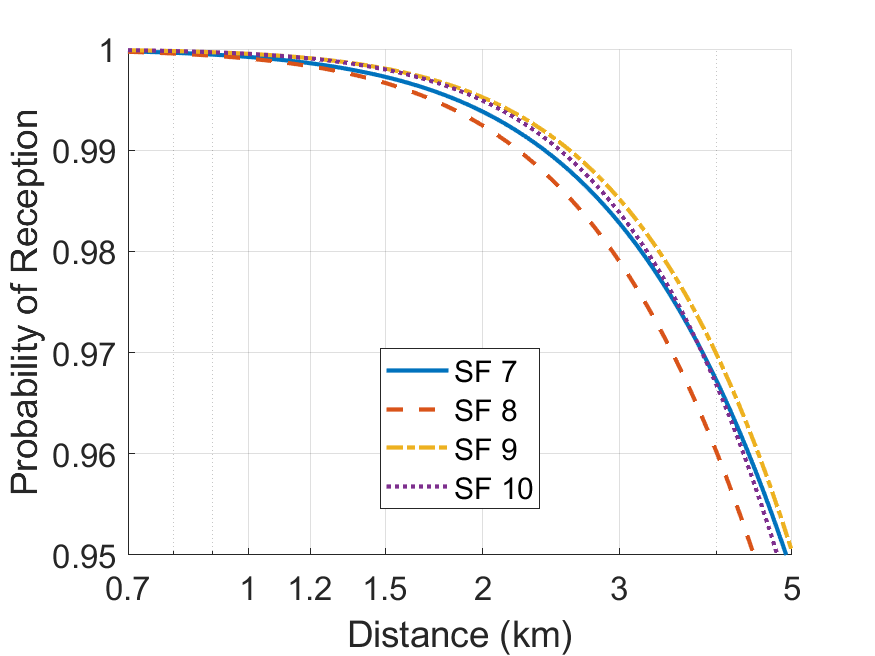}\label{fig:helikite_lw_PROB_RECEP_By_SF}}
    \hfill
    \subfloat[Boxplot of SNR vs distance by SF.]
        {\includegraphics[width=0.4\textwidth]{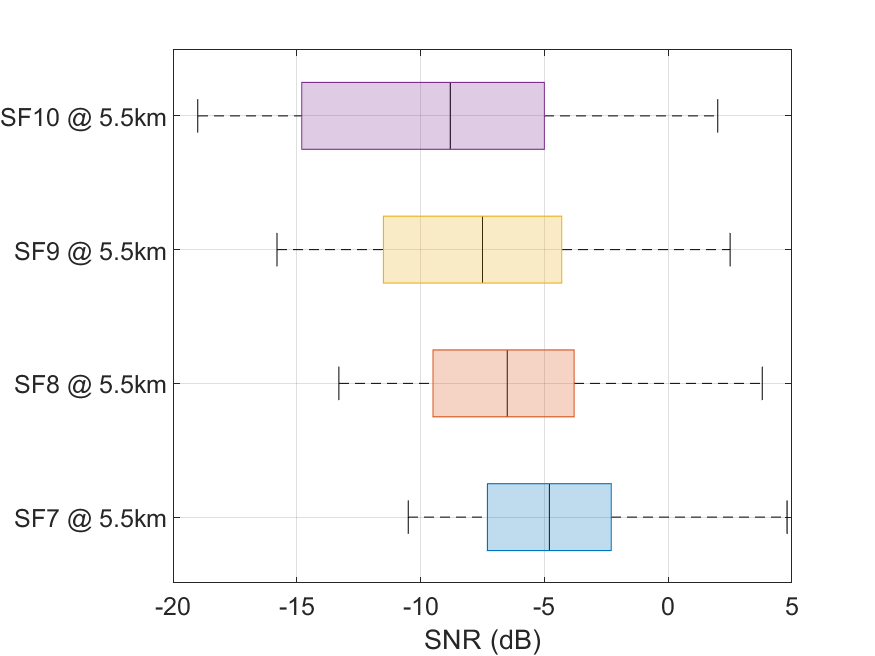}\label{fig:helikite_lw_BOX_SNR}}
    \caption{
        Helikite experiment results at Lake Wheeler Field (rural LOS): 
        (a) Trajectory and gateway locations with RSSI overlay,
        (b) RSSI CDF across gateways,
        (c) SNR CDF across gateways,
        (d) SNR versus distance by SF,
        (e) Probability of reception by SF and distance,
        (f) Boxplot of SNR versus distance by SF.
    }
    \label{fig:helikite_lw_all}
\end{figure*}


\subsubsection{Results and Analysis}

Figure~\ref{fig:car_all}a and Figure~\ref{fig:car_all}b present the empirical CDFs of RSSI and SNR values across the different gateways for the car-based experiment. Gateways CC2 and CC3, located in the urban Centennial Campus area, collected the highest number of packets, resulting in smoother distributions. LW1 and LW2, placed in the open rural field, show steeper CDF curves, reflecting more stable link quality. In contrast, LW4 and LW5 experienced sporadic reception, likely due to obstructed or edge-of-coverage positions. From Figure~\ref{fig:helikite_lw_all}a, they are both located close to trees that could obstruct the signals.

To better understand signal quality variation with respect to distance and SF, we evaluate the SNR trends from multiple perspectives. Figure~\ref{fig:car_all}c shows SNR as a function of distance, color-coded by SF. Threshold lines corresponding to the minimum demodulation SNR for each SF, based on Semtech specifications~\cite{SemtechSemtechSX1276}, are also included. While the data points overlap, the downward trend illustrates the expected SNR decay with distance. SF10, which tolerates the lowest SNR, dominates at longer distances.

To address the limitations of the scatter plot and better visualize SNR distributions per SF and distance bin, Figure~\ref{fig:car_all}d provides a boxplot representation. This plot highlights the statistical spread of SNR within predefined distance intervals, with separate distributions for each SF. It reveals that SF7 transmissions mostly occur at short ranges with limited SNR variation, while higher spreading factors, such as SF10, are more commonly used at longer distances and tend to show greater variation in SNR.

Figure~\ref{fig:car_all}e summarizes the probability of reception as a function of distance, estimated using a log-distance SNR fit for each SF. Each curve is generated by modeling the average SNR decay with distance and computing the reception probability in (\ref{eq:probReceive}) as the likelihood of exceeding the SNR threshold. The results show that SF7 maintains high reliability within short distances, but its reception drops off earlier than the others. SF8 and SF9 exhibit slightly extended coverage, while SF10 ensures the longest reliable range, as expected due to its lowest demodulation threshold.

Finally, Figure~\ref{fig:car_all}f presents the fitted log-distance path loss model for the car-based experiment. This plot compares measured path loss with the model curve, highlighting a close fit across a wide range of distances. The resulting model achieves a mean absolute error (MAE) of 8.06~dB, capturing the combined influence of rural and urban NLOS conditions observed during the drive.

\subsection{Helikite-Based Experiment}

The helikite experiments provide a unique perspective on signal propagation by using an aerial platform that maintains a stable altitude. Unlike mobile platforms such as vehicles or drones, the helikite hovers in place, moving only slightly due to wind. This setup reduces the variability introduced by changing positions and speeds, resulting in more consistent link measurements. Each campaign was conducted independently on different days—one at Lake Wheeler Field and the other at NC State Main Campus during the Packapalooza outreach event. In both cases, the helikite remained airborne for extended periods, allowing for the collection of a larger number of packets compared to other experiments.

\subsubsection{Lake Wheeler Field Results and Analysis}

\begin{figure*}[htbp]
    \centering
    \subfloat[Flight path and gateway locations during helikite experiment.]
        {\includegraphics[width=0.41\textwidth]{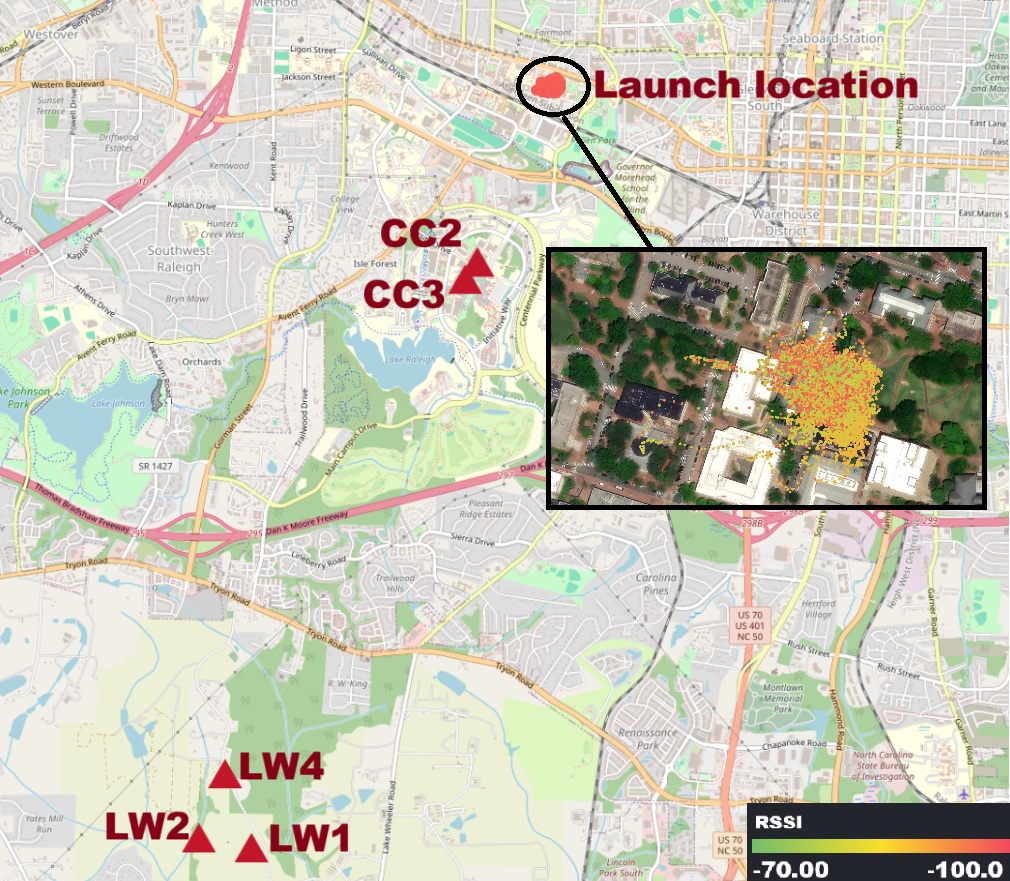}\label{fig:helikite_main_map}}
    \hfill
    \subfloat[CDF of RSSI across gateways. CC2 and CC3 are in urban environments, while others are in a rural suburban environment surrounded by trees.]
        {\includegraphics[width=0.41\textwidth]{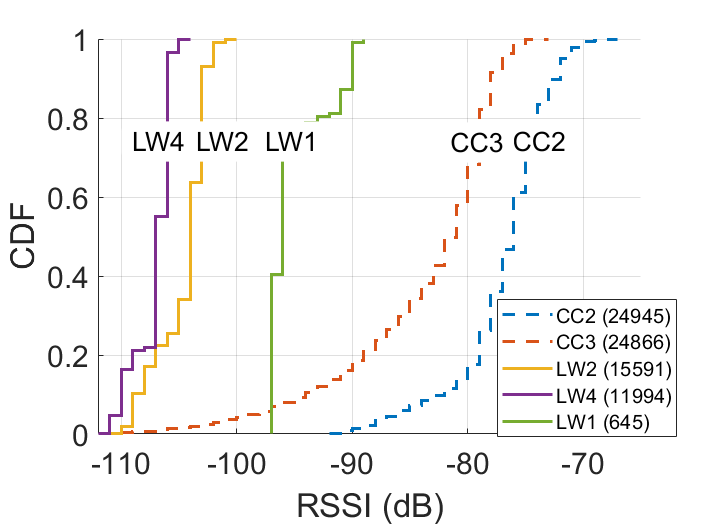}\label{fig:helikite_main_rssi_all}}
    \par\bigskip
    \subfloat[CDF of SNR across gateways. CC2 and CC3 are in urban environments, while others are in a rural suburban environment surrounded by trees.]
        {\includegraphics[width=0.41\textwidth]{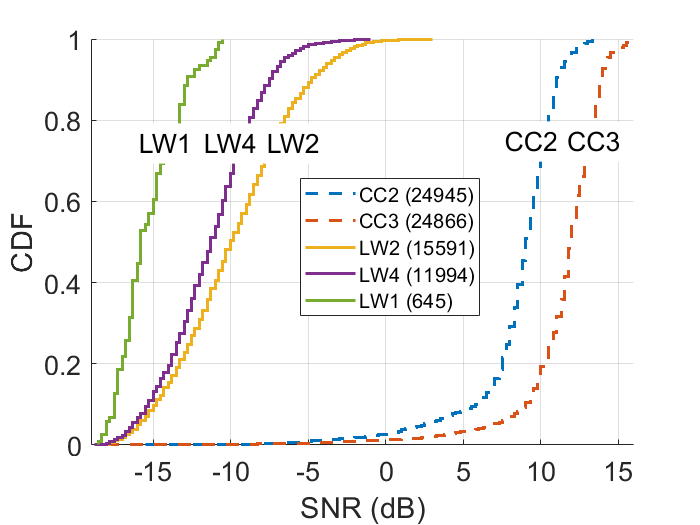}\label{fig:helikite_main_snr_all}}
    \hfill
    \subfloat[SNR vs distance by SF.]
        {\includegraphics[width=0.41\textwidth]{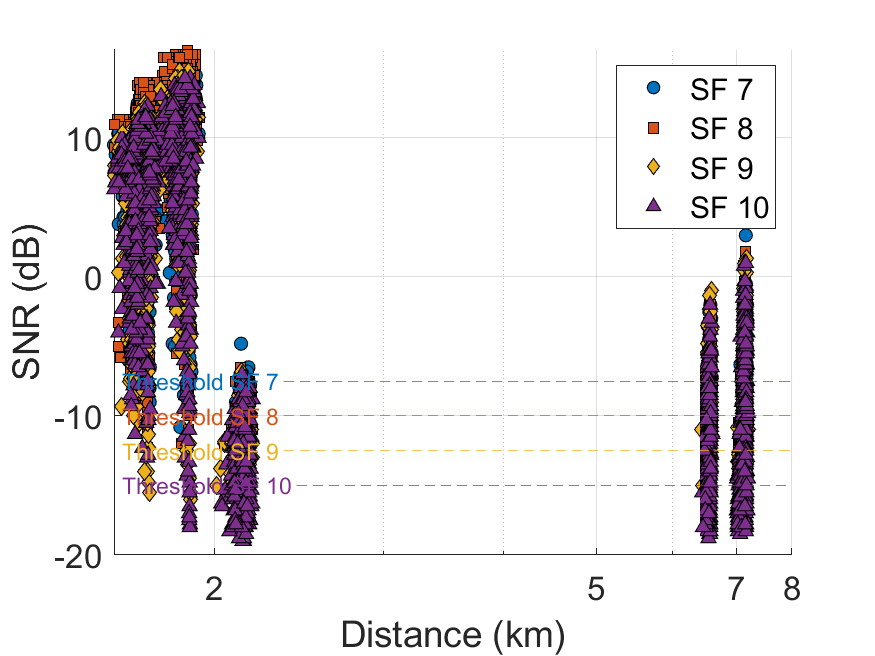}\label{fig:helikite_main_SNR_vs_Distance_By_SF}}
    \par\bigskip
    \subfloat[Probability of reception by SF.]
        {\includegraphics[width=0.41\textwidth]{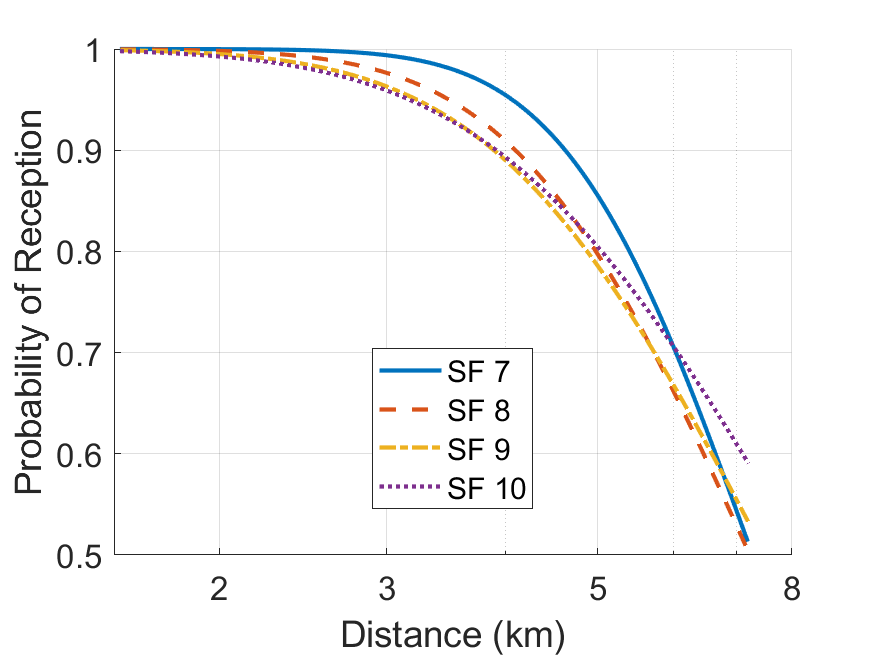}\label{fig:helikite_main_PROB_RECEP_By_SF}}
    \hfill
    \subfloat[Boxplot of SNR vs distance by SF.]
        {\includegraphics[width=0.41\textwidth]{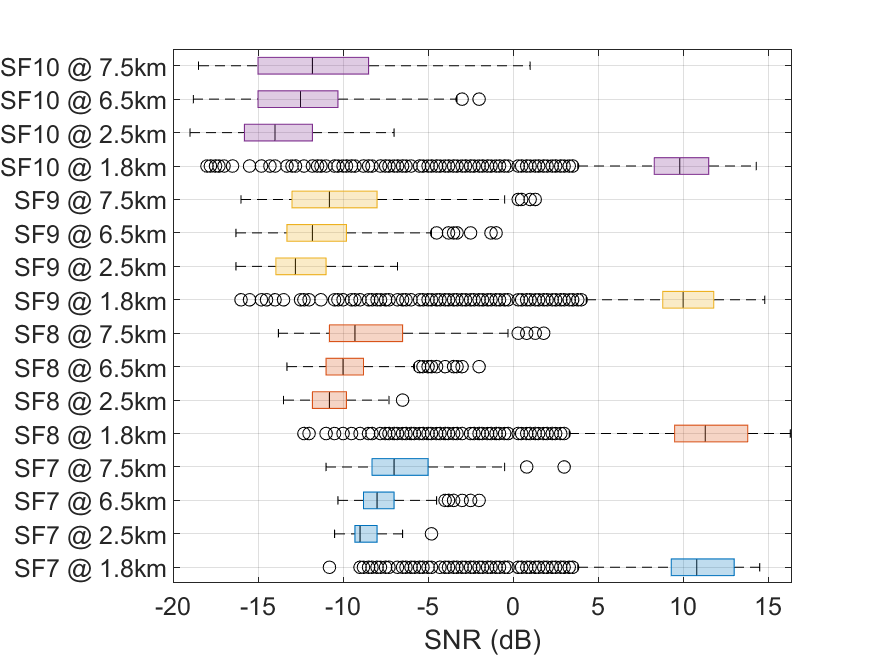}\label{fig:helikite_main_BOX_SNR}}
    \caption{
        Helikite experiment results (Main Campus, urban LOS): 
        (a) Flight path and gateway locations,
        (b) RSSI CDF across gateways,
        (c) SNR CDF across gateways,
        (d) SNR vs. distance by SF,
        (e) Probability of reception by SF and distance,
        (f) SNR boxplot vs. distance by SF.
    }
    \label{fig:helikite_main_campus_all}
\end{figure*}

The helikite experiment at Lake Wheeler Field, shown in Figure~\ref{fig:helikite_lw_all}a, was conducted under rural, line-of-sight (LOS) conditions. The platform remained tethered at a constant altitude of 150~meters, enabling precise and stable transmission paths to the receiving gateways. Only gateways LW1, LW2, LW4, and CC2 were operational during the test.

Figures~\ref{fig:helikite_lw_all} b and~\ref{fig:helikite_lw_all}c show the cumulative distributions of RSSI and SNR, respectively, for the four active gateways. The field-based receivers (LW1, LW2, LW4) recorded stronger signal levels compared to CC2, which is located farther away. The CDF curves for these gateways are tightly grouped, reflecting consistent reception from the nearly stationary airborne platform. In contrast, CC2 shows a distribution shifted toward lower RSSI and SNR values, highlighting its relatively disadvantaged link quality.

Figure~\ref{fig:helikite_lw_all}d presents SNR versus distance grouped by SF. The SNR levels remain high across all SFs, with only SF10 showing some points below its reception threshold. This confirms that LOS conditions provided fa

\subsubsection{Main Campus Results and Analysis}

The helikite experiment conducted over NC State’s Main Campus involved stationary transmissions from a 150~meter altitude, enabling a clear line-of-sight (LOS) to most of the surrounding gateways. Figures~\ref{fig:helikite_main_campus_all}b and~\ref{fig:helikite_main_campus_all}c display the cumulative distribution functions (CDFs) of RSSI and SNR across gateways. CC2, the closest gateway, shows the strongest and most stable signal quality, followed by LW4 and LW1. The steeper slope in CC2’s CDF confirms lower variability in both RSSI and SNR, which is expected due to its proximity and unobstructed view to the helikite.

Figure~\ref{fig:helikite_main_campus_all}e shows how SNR varies with distance for each spreading factor (SF7–SF10), alongside their respective theoretical demodulation thresholds. As expected, higher SFs such as SF10 are predominantly used at greater distances, while lower SFs are more common at shorter ranges. The data clusters near 2.5~km and 7.5~km correspond to groups of gateways, illustrating the segmented reception regions in the experiment.

Figure~\ref{fig:helikite_main_campus_all}e plots the estimated probability of successful reception as a function of distance, derived from a linear model fit of SNR versus $\log_{10}$(distance). The results confirm that SF10 offers the most extended effective range, maintaining a reception probability above 0.9 up to 7~km, while SF7 performance declines more rapidly beyond 5~km. This reinforces the advantage of higher SFs for extended coverage in urban environments.

Finally, Figure~\ref{fig:helikite_main_campus_all}f shows a boxplot of SNR values grouped by SF and distance range. At short distances (around 1.8~km), all SFs from 7 to 10 are present. In this range, SF7 and SF8 exhibit high median SNR values and tight distributions, indicating strong and stable signal quality. SF9 and SF10 are less common here, as they are typically reserved for longer distances. As distance increases, lower SFs such as SF7 and SF8 appear less frequently, while SF9 and SF10 dominate. For instance, SF10 is mostly observed at 6.5~km and 7.5~km, where SNR is lower.

The box widths further reveal link stability. SNR values are more consistent at short ranges, while the spread widens significantly beyond 6~km. This suggests increased signal variation at long range, potentially caused by small-scale obstacles or elevation differences. This analysis confirms that LoRaWAN dynamically uses higher SFs to maintain connectivity at longer distances. The helikite platform demonstrated stable and reliable performance, maintaining acceptable SNR levels across a wide urban area.

\subsection{Drone-Based Experiment}
The drone experiment offers a mobile aerial transmission scenario at a fixed altitude of approximately 50~meters. Unlike the helikite, which remained airborne for extended durations over a small area, the drone followed a pre-planned flight path over Lake Wheeler Field. This resulted in a shorter experiment duration and a lower number of received packets. The reduced sampling density offers a contrasting perspective on LoRaWAN performance under dynamic aerial movement. Due to the relatively low altitude and the longer distance to some gateways, the drone operated under NLOS conditions with respect to the Centennial Campus, where terrain elevation, buildings, and vegetation obstructed the signal path. No packets were received at the Centennial Campus gateways (CC2 and CC3).

\subsubsection{Results and Analysis.} 
\begin{figure*}[htbp]
    \centering
    \subfloat[Flight path and gateway locations during drone experiment]
        {\includegraphics[width=0.42\textwidth]{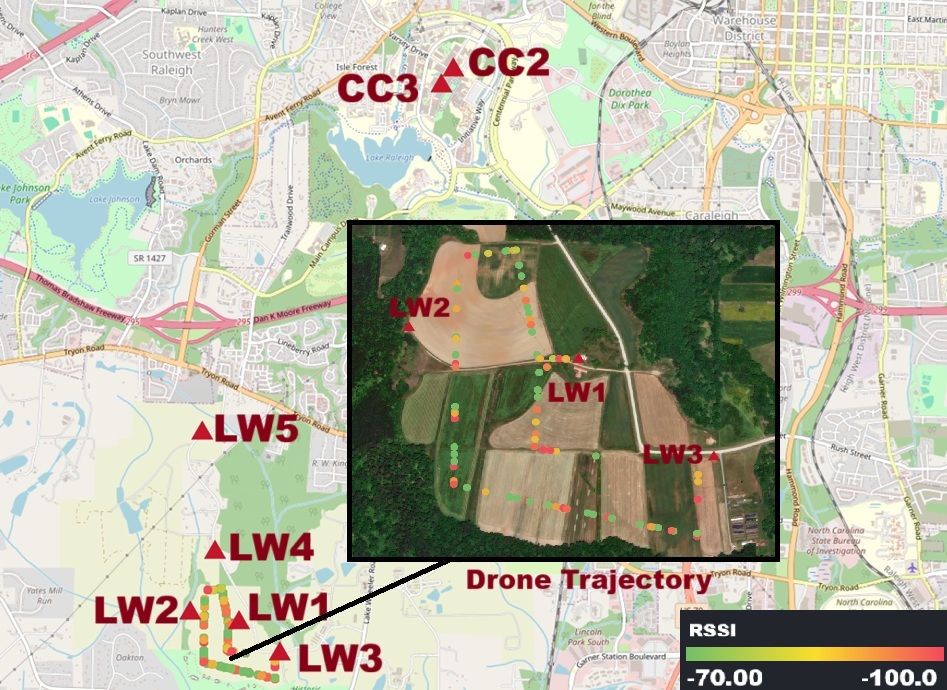}\label{fig:drone_lw_map}}
    \hfill
    \subfloat[CDF of RSSI across gateways. CC2 is in urban environment, while others are in a rural suburban environment surrounded by trees.]
        {\includegraphics[width=0.45\textwidth]{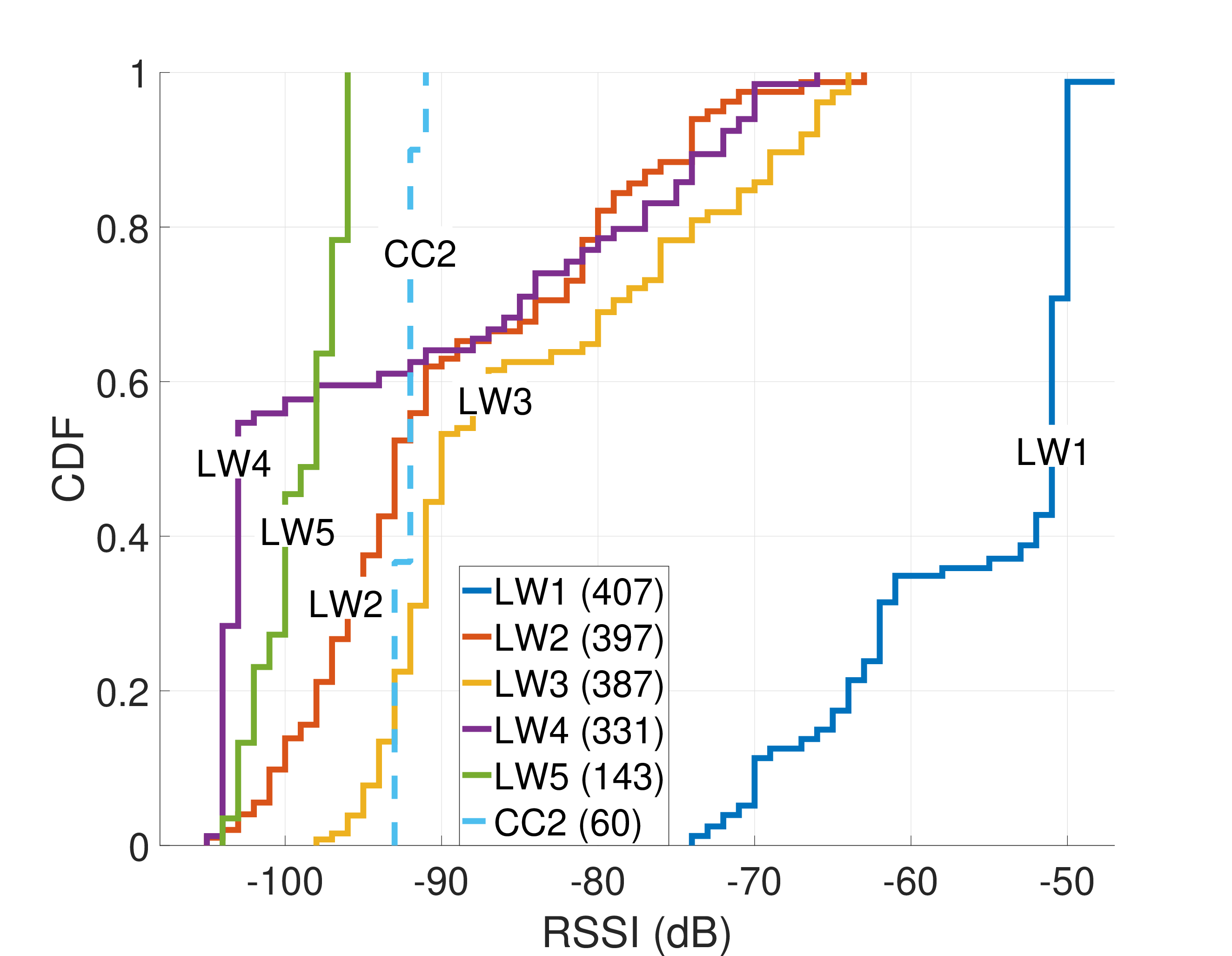}\label{fig:drone_lw_rssi_cdf}}
    \par\bigskip
    \subfloat[CDF of SNR across gateways. CC2 is in urban environment, while others are in a rural suburban environment surrounded by trees.]
        {\includegraphics[width=0.45\textwidth]{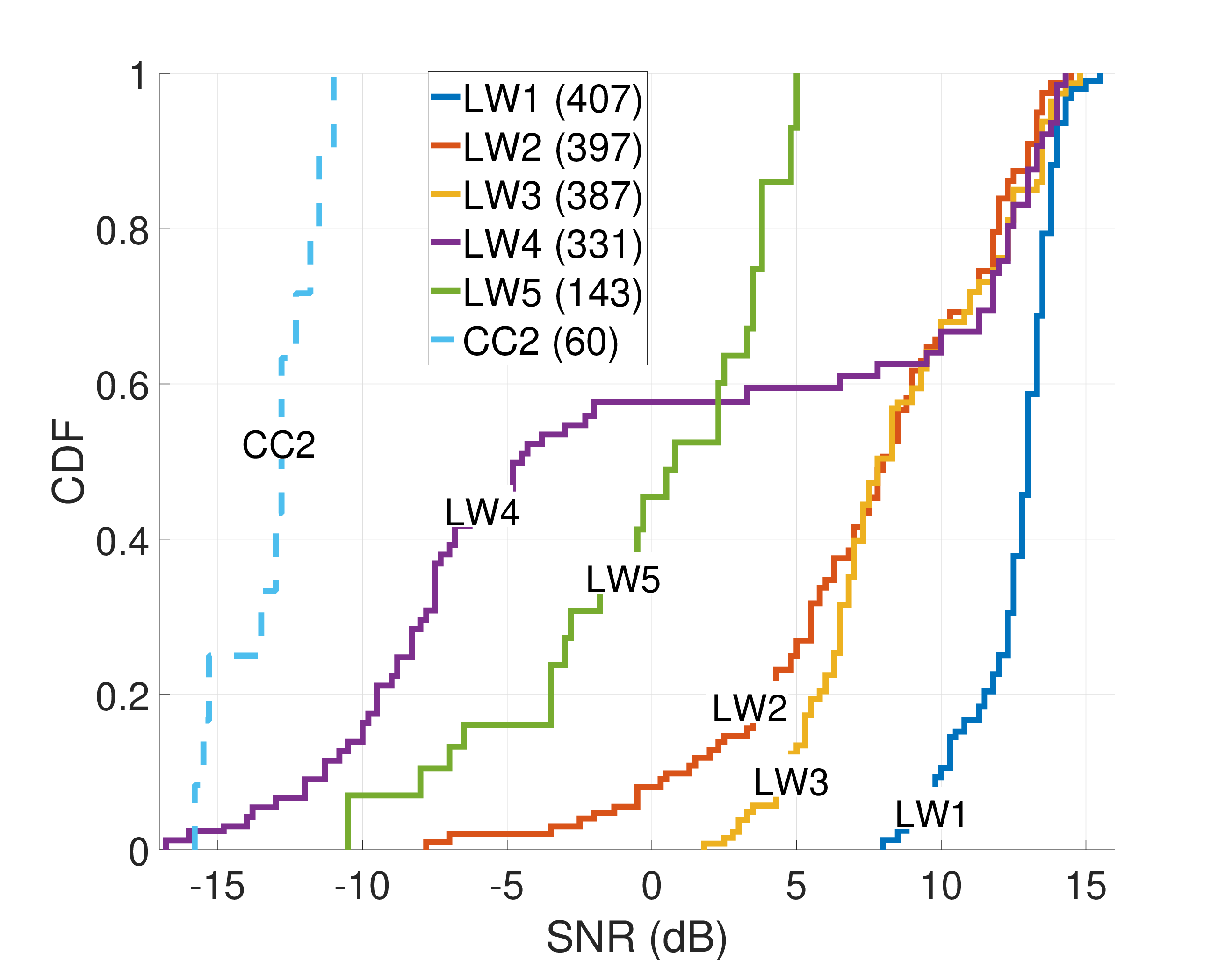}\label{fig:drone_lw_snr_cdf}}
    \hfill
    \subfloat[SNR vs distance by SF.]
        {\includegraphics[width=0.45\textwidth]{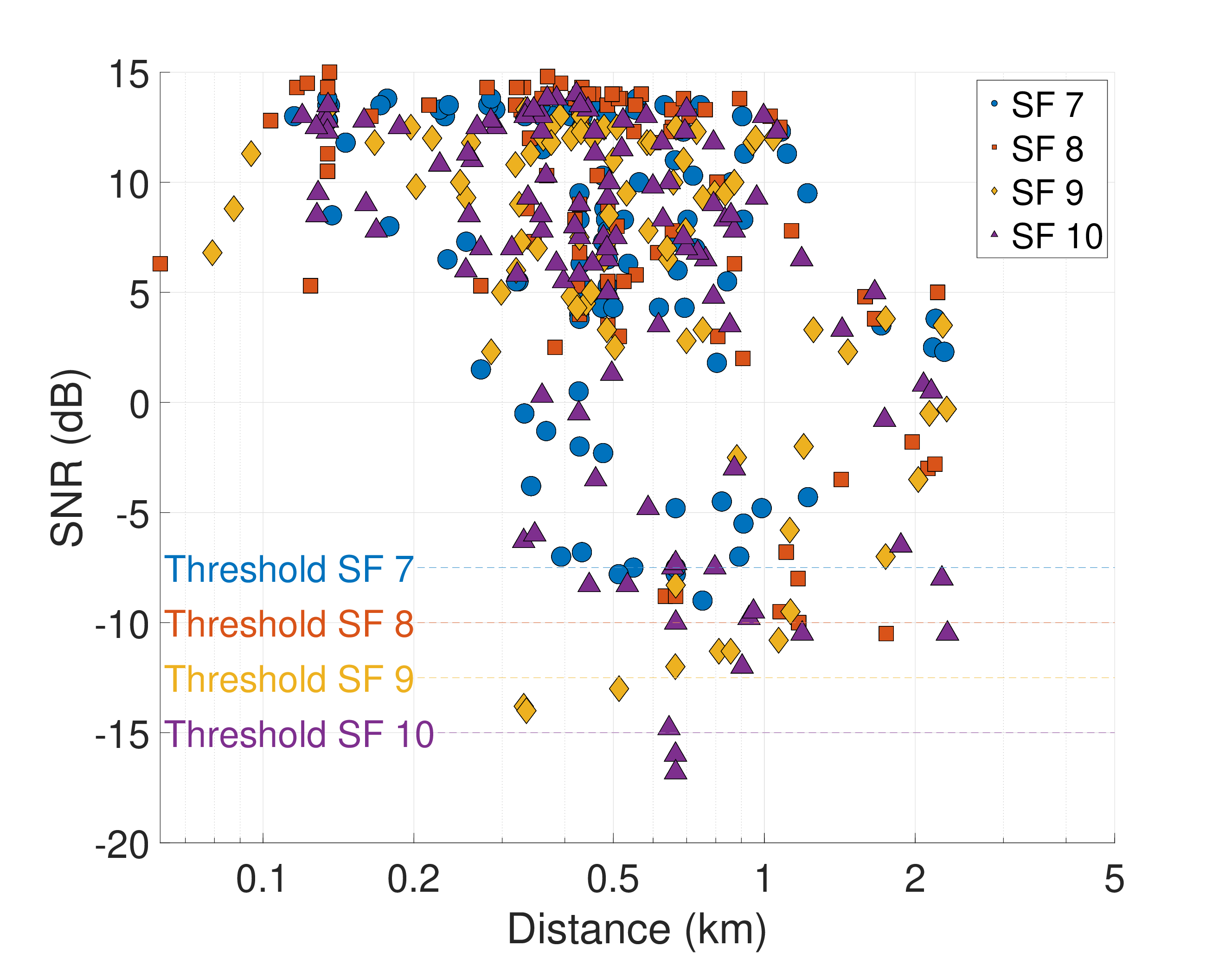}\label{fig:drone_lw_snr_vs_distance_by_sf}}
    \par\bigskip
    \subfloat[Probability of reception by SF.]
        {\includegraphics[width=0.45\textwidth]{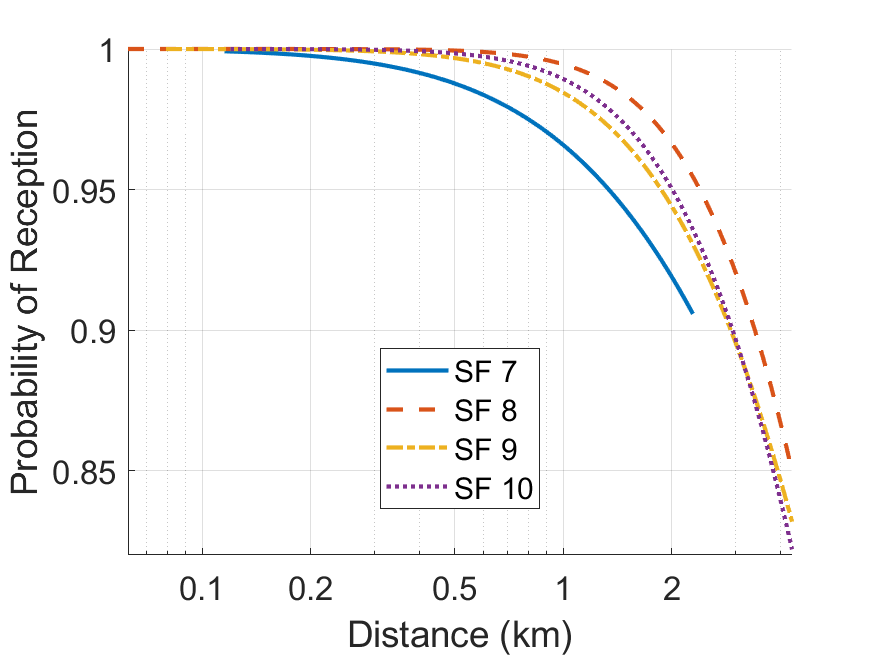}\label{fig:drone_lw_prob_recep_by_sf}}
    \hfill
    \subfloat[Boxplot of SNR vs distance by SF.]
        {\includegraphics[width=0.43\textwidth]{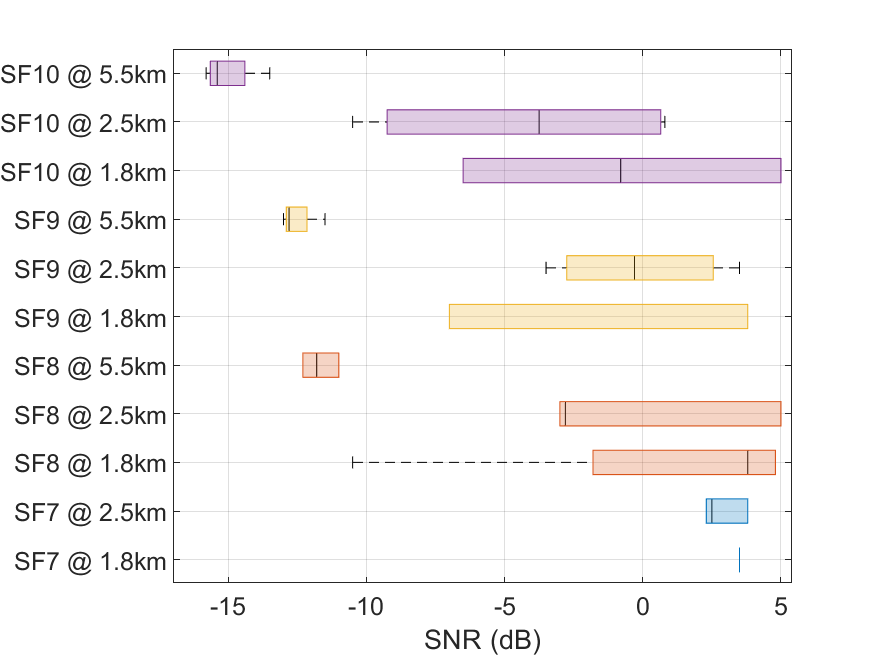}\label{fig:drone_lw_box_snr}}
    \caption{
        Drone experiment results at Lake Wheeler Field: 
        (a) Flight path and gateway locations,
        (b) RSSI CDF across gateways,
        (c) SNR CDF across gateways,
        (d) SNR vs. distance by SF,
        (e) Probability of reception by SF and distance,
        (f) SNR boxplot vs. distance by SF.
    }
    \label{fig:drone_lw_all}
\end{figure*}

Figures~\ref{fig:drone_lw_all}b and~\ref{fig:drone_lw_all}c show the cumulative distribution functions (CDFs) of RSSI and SNR per gateway. Compared to the helikite results, these distributions are more irregular and spread out, reflecting the drone's changing heading. LW1 and LW2 recorded the most packets with relatively stronger signal levels, while gateways such as LW3 and LW5 received fewer packets, indicating less favorable link conditions or intermittent LOS.

Figure~\ref{fig:drone_lw_all}d plots SNR as a function of distance and SF. The plot includes the demodulation thresholds for each SF, which serve as reception boundaries. A significant portion of the SF9 and SF10 data points fall near or below their respective thresholds, especially at distances greater than 1~km, suggesting sensitivity-limited reception under drone mobility.

The reception probability curves in Figure~\ref{fig:drone_lw_all}e confirm this trend. SF7 and SF8 maintain higher reception probabilities up to around 1.5~km, whereas SF9 and SF10 degrade more quickly with distance. These results show that under mobile aerial conditions, even moderate altitude cannot fully compensate for longer distances and fading effects.

The boxplot in Figure~\ref{fig:drone_lw_all}f provides a detailed view of SNR variation across different SFs and distance intervals. Compared to the helikite case, the boxplots here are noticeably wider and more dispersed, particularly for SF9 and SF10. This suggests greater fluctuation in received signal quality during the drone flight. The wider spread likely reflects the effects of aerial mobility, including heading changes, brief signal blockages, and rapid transitions between LOS and NLOS. The SNR values for SF10 at 5.5~km, for instance, span a broad range, with some falling below the demodulation threshold, indicating marginal link conditions at that distance.

\section{Conclusion}\label{sec:conclusion}

This paper evaluated the propagation performance of a LoRaWAN network deployed at NC State University's Centennial Campus and Lake Wheeler Field through three distinct measurement campaigns: a car-based ground experiment, a helikite-based aerial experiment, and a drone-based mobile aerial experiment. Each platform offered insights into LoRaWAN signal behavior under varying conditions influenced by altitude, mobility, and LOS.

The helikite experiments at Lake Wheeler and Main Campus recorded the highest number of received packets due to their extended duration and elevated, mostly stationary position. With the transmitter at approximately 150~meters, link quality remained consistently strong across multiple gateways, demonstrating high SNR and reliable reception even over long distances.

The drone experiment, flown at 50~meters, collected fewer packets and exhibited greater signal variation. Its lower altitude and increased distance to urban gateways resulted in limited coverage and no reception at Centennial Campus, consistent with partial NLOS conditions.

The car-based experiment captured the effects of terrain, vegetation, and urban structures on LoRaWAN performance under ground-level mobility. It enabled the application of the log-distance path loss model to characterize signal behavior in mixed rural and urban environments, providing clear insight into distance-dependent attenuation.

We further analyzed SNR as a function of distance, modeled reception probability per spreading factor (SF), and visualized SNR distributions using boxplots grouped by SF and distance. Results show that higher SFs are more prevalent at longer distances or under reduced link quality, reflecting the adaptive nature of LoRaWAN.

Overall, the results suggest that high-altitude platforms such as the helikite offer reliable, wide-area coverage, while mobile platforms highlight performance variability under dynamic or constrained conditions. These findings underscore the importance of empirical measurements in deployment planning and support the use of measurement-driven models for optimizing LoRaWAN networks in diverse environments.

\section*{Acknowledgements}\label{sec:acknowledgements}
This work was supported by the NSF PAWR program under grant number CNS-1939334. 
The datasets and post-processing scripts used to generate the results in this paper 
are available at \url{https://aerpaw.org/experiments/datasets/}. 
The authors would like to thank Thomas Zajkowski, Flight Operations Manager at AERPAW, 
and the entire AERPAW team for their support and contributions. 
The authors also acknowledge the use of OpenAI’s ChatGPT (GPT-5) to assist with 
language editing, text organization, and overall writing clarity during the preparation 
of this manuscript.

\bibliography{references_manual}

\thebiography

\begin{biographywithpic}
{Sergio Vargas Villar}{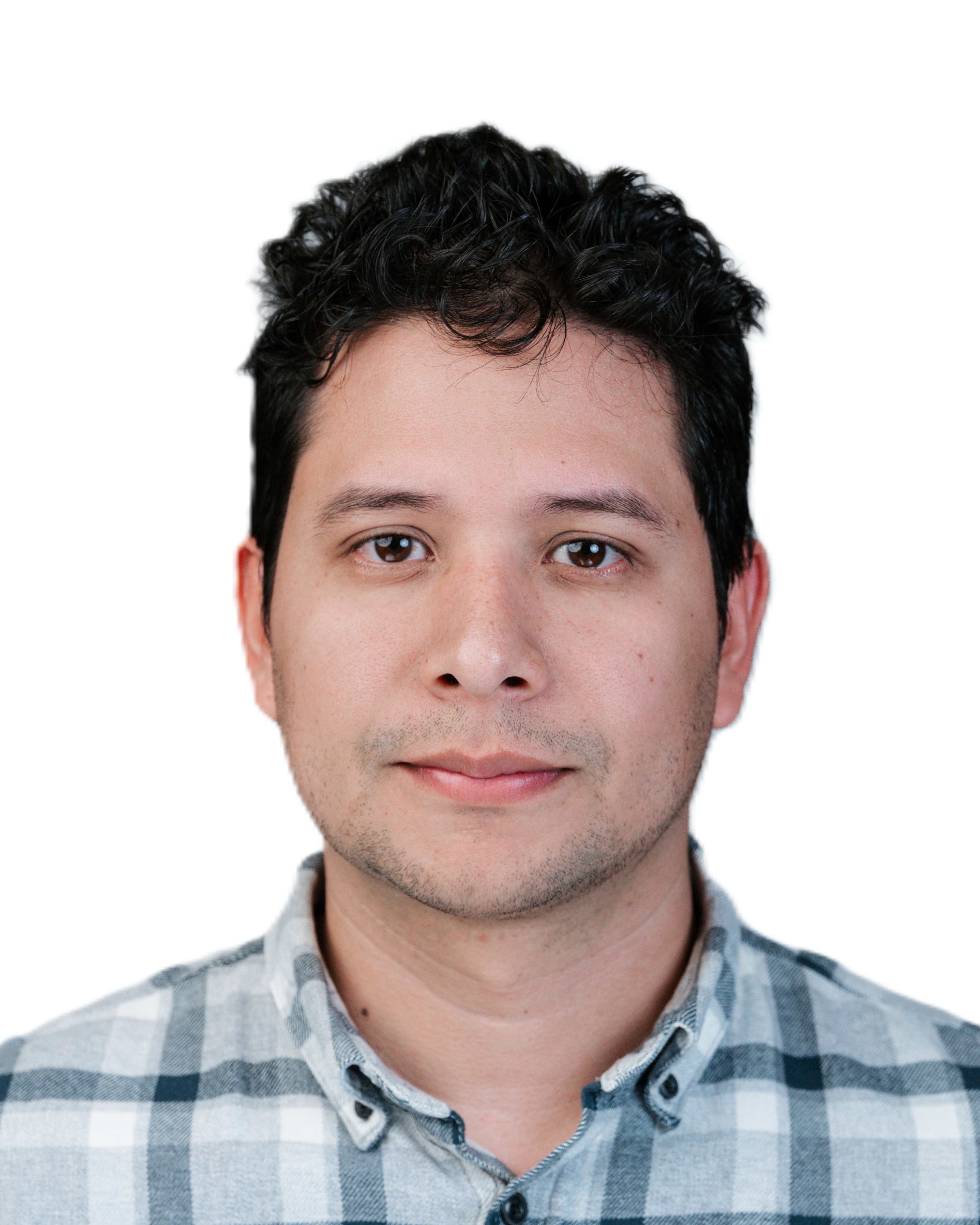} 
\itshape received the B.S. in Mechatronics Engineering from the Bolivian Catholic University (2015), the M.S. in Industrial Automation and Robotics from the Polytechnic University of Catalonia (2019), and the M.Sc. in Electrical Engineering from North Carolina State University (2025). He has 3 years of experience as a hardware engineer at Bettair Cities in Spain, where he developed IoT hardware for air quality monitoring, and 3 years as a full-stack developer at Innovent Tech, working on RFID and Industry 4.0 systems. He also served as an adjunct professor at the Bolivian Catholic University, teaching Robotics and Communication systems. At NC State, he contributed to LoRaWAN and 5G integration through the AERPAW project. He is currently an Electrical Design Engineer at Zoetis Veterinary Medicine Research and Development (VMRD).
\end{biographywithpic}

\begin{biographywithpic}
{Simran Singh}{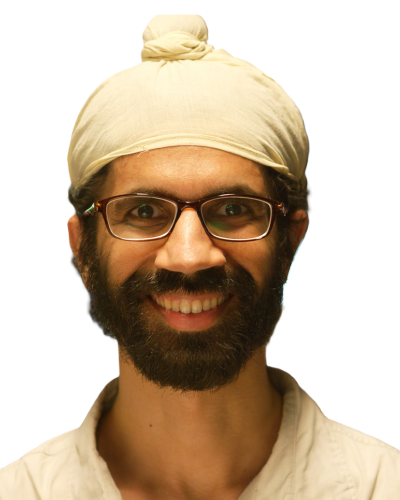} 
\itshape received his B.Tech. degree in Electronics from Veermata Jijabai Technological University Institute, Mumbai, India, in 2011. He worked as a software engineer for four years in the computer-aided design (CAD) domain, building both native and full-stack web applications. He obtained his M.S. in Computer Networks and Ph.D.
in Electrical Engineering from NCSU in 2022. He is currently a postdoctoral researcher at NCSU. His research interests include wireless communications for unmanned aerial vehicles.
\end{biographywithpic}

\begin{biographywithpic}
{Ozgur Ozdemir}{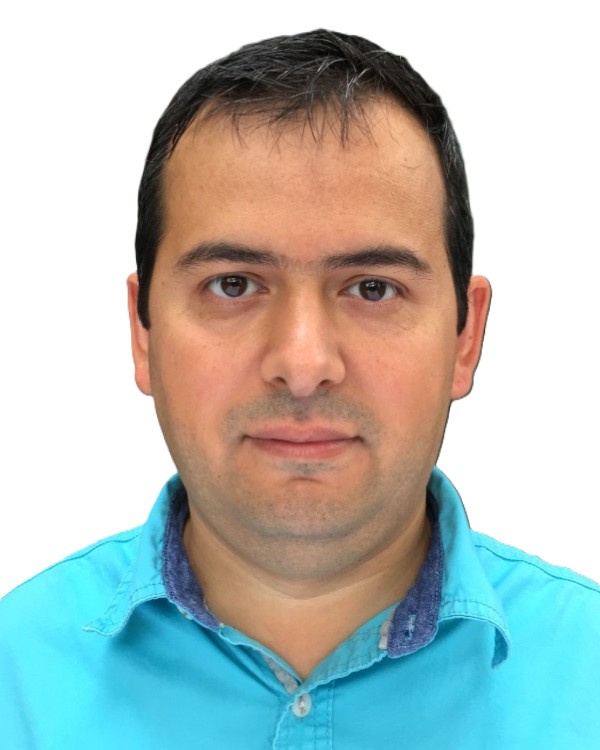} 
 received the BS degree in Electrical and Electronics Engineering from Bogazici University, Istanbul, Turkey, in 1999 and the MS and Ph.D. degrees in Electrical Engineering from The University of Texas at Dallas, Richardson, TX, USA, in 2002 and 2007, respectively. From 2007 to 2016, he was an Assistant Professor at Fatih University, Turkey, and worked as a Postdoctoral Scholar at Qatar University for 3.5 years. He joined the Department of Electrical and Computer Engineering at NC State as a visiting research scholar in 2017 and is now serving as an Associate Research Professor. His research interests include software-defined radios, channel sounding for mmWave systems, wireless testbeds, digital compensation of radio-frequency impairments, and opportunistic approaches in wireless systems. He is serving as the lead for supporting field experimentation with wireless technologies and drones for the NSF AERPAW platform at NC State.
\end{biographywithpic}

\sloppy
\begin{biographywithpic}
{Prof. Mihail L. Sichitiu}{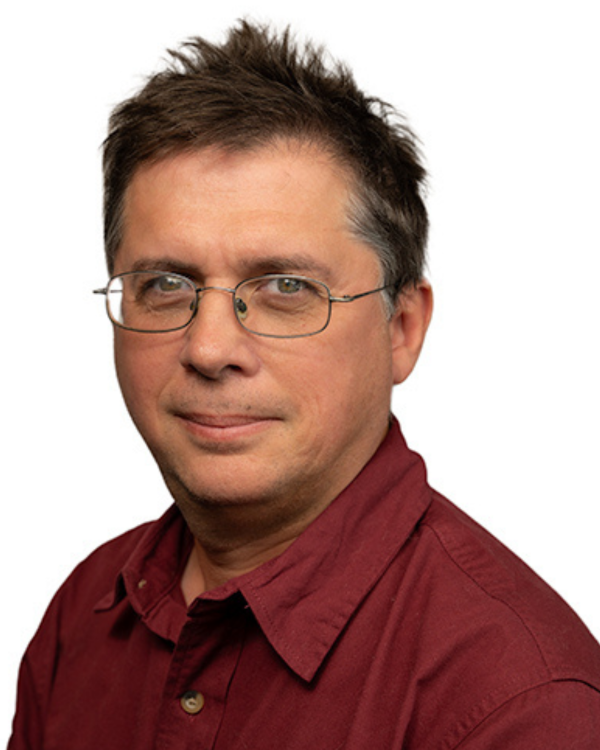} 
\itshape earned his Ph.D. degree in Electrical Engineering from the University of Notre Dame in 2001. His current research interests include wireless networks and communications for UAVs. In these systems, he is studying problems related to localization, time synchronization, emulation, routing, fairness, and modeling. He is teaching wireless networking and UAV courses. He is a professor in the department of Electrical and Computer Engineering at NCSU.
\end{biographywithpic}

\sloppy
\begin{biographywithpic}{Prof. \.{I}smail G\"{u}ven\c{c}}{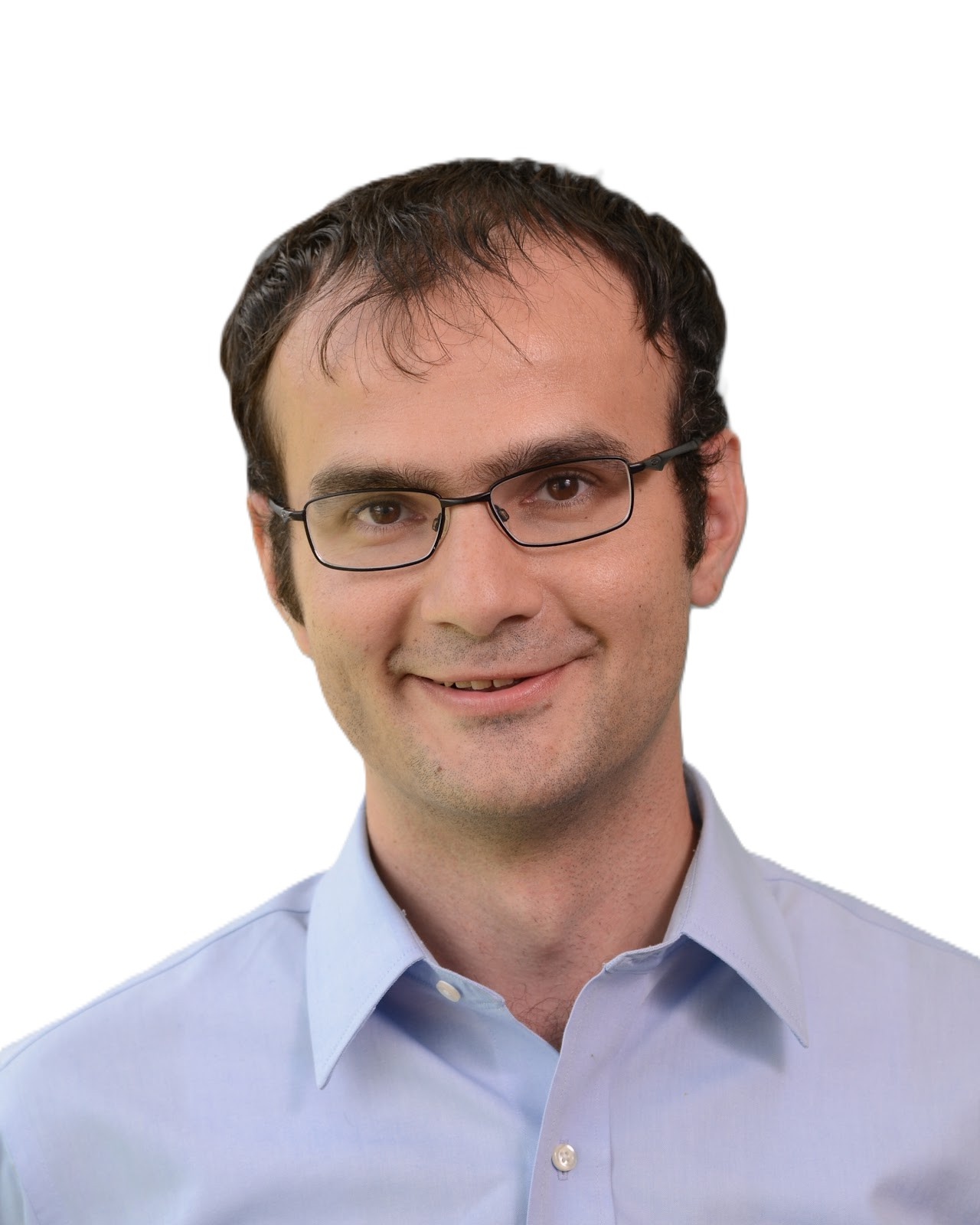}
\itshape is a Professor at the Department of Electrical and Computer Engineering at NC State University. His recent research interests include 5G/6G wireless networks, UAV communications, millimeter/terahertz communications, and heterogeneous networks. He has published more than 300 conference/journal papers and book chapters, several standardization contributions, four books, and over 30 U.S. patents. Dr. Guvenc is the PI and the director for the NSF AERPAW project and a site director for the NSF BWAC I/UCRC center. He is an IEEE Fellow, a senior member of the National Academy of Inventors, and a recipient of several awards, including NC State University Alcoa Distinguished Engineering Research Award (2023), Faculty Scholar Award (2021), R. Ray Bennett Faculty Fellow Award (2019), FIU COE Faculty Research Award (2016), NSF CAREER Award (2015), Ralph E. Powe Junior Faculty Award (2014), and USF Outstanding Dissertation Award (2006).
\end{biographywithpic}

\end{document}